\documentclass[pra,aps,twocolumn,amsmath,amssymb,floatfix,pra,reprint,footinbib,superscriptaddress,longbibliography]{revtex4-2}
\usepackage{graphics}      
\usepackage{graphicx}      
\usepackage{longtable}     
\usepackage{url}           
\usepackage{bm}    
\usepackage{braket}        
\usepackage{xcolor}
\usepackage{float}  
\usepackage{natbib}
\usepackage{graphicx}
\usepackage{comment}
\usepackage{siunitx}
\usepackage{appendix}
\usepackage{amsmath,amssymb,epsfig,color}

\usepackage[most]{tcolorbox}

\newcommand{\nn}{\nonumber}
\newcommand{\nl}{\nonumber \\}
\newcommand{\be}{\begin{equation}}
\newcommand{\ee}{\end{equation}}
\newcommand{\bea}{\begin{eqnarray}}
\newcommand{\eea}{\end{eqnarray}}

\usepackage{amssymb}
\usepackage{mathrsfs}
\usepackage{amsfonts}
\usepackage{graphicx}
\usepackage{amsmath}
\usepackage{dcolumn}
\usepackage{color}
\usepackage{subfigure}
\usepackage{epsfig}
\usepackage{soul}
\usepackage[colorlinks,linkcolor=blue,citecolor=blue,hyperindex,bookmarks=false,pdfstartview=FitH]{hyperref}

\providecommand{\U}[1]{\protect\rule{.1in}{.1in}}

\newcommand{\figpanel}[2]{\hyperref[#1]{\ref*{#1}(#2)}}

\begin{document}

\title { Engineering Bosonic Codes with Quantum Lattice Gates }

\author{Lingzhen Guo}
\affiliation{Center for Joint Quantum Studies and Department of Physics, School of Science, Tianjin University, Tianjin 300072, China}

\author{Tangyou Huang}
\email{tangyou@chalmers.se}
\affiliation{Department of Microtechnology and Nanoscience (MC2), Chalmers University of Technology, SE-41296 G\"oteborg, Sweden}


\author{Lei Du}
\email{lei.du@chalmers.se}
\affiliation{Department of Microtechnology and Nanoscience (MC2), Chalmers University of Technology, SE-41296 G\"oteborg, Sweden}

\begin{abstract}
Bosonic codes offer a hardware-efficient approach to encoding and protecting quantum information with a single continuous-variable bosonic system. In this paper, we introduce a new universal quantum gate set composed of only one type of gate element, which we call the \textit{quantum lattice gate},
 to engineer bosonic code states for fault-tolerant quantum computing. We develop a systematic framework for code state engineering based on the Floquet Hamiltonian engineering, where the target Hamiltonian is constructed directly from the given target state(s). 
We apply our method to three basic code state engineering processes, including single code state preparation, code space embedding and code space transformation. We explore the application of our method to automatic quantum error correction against single-photon loss with four-legged cat codes.
Our proposal is particularly well-suited for superconducting circuit architectures with Josephson junctions, where the full nonlinearity of Josephson junction potential is harnessed as a quantum resource and the quantum lattice gate can be implemented on a sub-nanosecond timescale.
\end{abstract}

\date{\today}

\maketitle

\section{~Introduction}

Quantum computers take advantages of quantum coherence and entanglement that do not exist in classical world to process information. A range of important problems such as factorizing large integers \cite{shor1997siam,ekert1996rmp} and simulating quantum many-body dynamics \cite{smith2019npj,fauseweh2024nc} are promised to be solved by quantum computers exponentially faster than classical computers \cite{chuang2010book}. The basic unit of a quantum computer to encode information is the \textit{quantum bit} (qubit) made from \textit{discrete-variable} (DV) system such as spins and quantized levels. However, due to the ubiquitous noises in the environment, quantum states are fragile and the encoded information in qubits can be inevitably deteriorated and lost \cite{preskill2018quatum}.
For practical quantum computing, it is crucial to implement \textit{quantum error correction} (QEC) to protect against unwanted and uncontrolled errors, which remains one of the most challenging and urgent goals for building a fault-tolerant quantum computer~\cite{preskill1998book,devoret2013science,Ofek2016nature}. Representative QEC codes with DV qubits towards practical large-scale quantum computation include \textit{stabilizer codes} \cite{gottesman1997stabilizer} and \textit{surface codes} \cite{Dennis2002jmp,kitaev2003aop,fowler2012pra}, which encode a \textit{logical qubit} in a subspace of multiple \textit{physical qubits} so that different error events lead to distinguishable detectable syndromes, thereby facilitating the recovery of corrupted quantum states. 
However, such DV-based QEC schemes typically consume substantial physical resource, as encoding a single logical qubit often demands a large number of physical qubits. Furthermore, logical gate operations in these schemes are complicated, as multiple physical qubits have to be manipulated simultaneously. Scaling up physical qubits to build a fault-tolerant quantum computer remains an exceptionally challenging task due to these significant demands on physical qubits and operation complexity~\cite{corcoles2020ieee,Gidney2021howtofactorbit}. 

The core idea of QEC lies in protecting the encoded quantum states by introducing a redundant Hilbert space for error detection and correction.  A single harmonic oscillator already provides an infinitely large Hilbert space that can be partitioned into  a logical subspace and error subspace \cite{chuang1997pra,braunstein1998prl,Gottesman2001pra,Cochrane1999pra,Michael2016prx}. \textit{Continuous-variable} (CV) systems, such as harmonic oscillators or other bosonic systems, naturally provide an infinite-dimensional Hilbert space that can be partitioned into logical and error subspaces, making them an attractive alternative for fault-tolerant universal quantum computation\cite{chuang1997pra,braunstein1998prl,Gottesman2001pra,Cochrane1999pra,Michael2016prx}. Unlike DV systems, CV systems leverage the continuous nature of their state space, which inherently offers redundancy without the need of a large number of physical qubits~\cite{Braunstein2005RMP,weebrook2012rmp}.%
It has been proved that a single nonlinear term in addition to linear Gaussian operations are possible to realize universal single-mode CV quantum computation \cite{lloyd1999prl}, e.g., a common universal gate set includes the cubic phase gate, displacement gate, squeezing gate, and Fourier gate (phase rotation). In the superconducting circuit experiments, the cubic  phase gate has been proposed and realized with an on-chip planar resonator terminated by a \textit{superconducting nonlinear asymmetric inductive element} (SNAIL) \cite{hillmann2020prl,eriksson2024nc}. 
Another popular universal set for single-mode bosonic quantum computation is composed of a displacement gate and a \textit{selective number-dependent arbitrary phase} (SNAP) gate, which imparts an arbitrary phase to every Fock number state \cite{Krastanov2015pra} using an off-resonantly coupled qubit \cite{heeres2015prl}.
Universal CV quantum computation for multiple bosonic modes can be implemented by adding a simple two-mode linear operation \cite{sefi2011prl,niklas2022prr}.

However, the above universal gate operations for CV quantum computation are not guaranteed to be fault tolerant due to the intrinsic continuous error channels of CV mode quadratures. To achieve practical fault-tolerant quantum computation with CV systems, it is necessary to embed a finite-dimensional code space into the infinite-dimensional CV Hilbert space, similar to QEC using DV systems~\cite{Travaglione2002pra}. Since the most common and practically relevant Gaussian CV errors generally cannot be suppressed solely based upon Gaussian states and Gaussian operations \cite{niset2009prl}, the discretized fault-tolerant code states must be non-Gaussian states. But not all the non-Gaussian states are suitable for fault-tolerant quantum computing. Depending on the error set relevant in experiments, the construct discretized non-Gaussian states have to satisfy the so-called  Knill-Laflamme condition \cite{knill1997pra,chuang2010book}, and are known as \textit{bosonic QEC codes}, or simply \textit{bosonic codes}~\cite{weizhou2021fr,wenlong2021sb}. According to the symmetries in phase space, the bosonic codes can be broadly classified into the \textit{translational codes} and \textit{rotational codes} \cite{Arne2020prx}. A prominent example of translational codes is the \textit{Gottesman-Kitaev-Preskill} (GKP) code~\cite{Gottesman2001pra}, while rotational bosonic codes include \textit{cat code}~\cite{Cochrane1999pra,Leghtas2013prl} and \textit{binomial code}~\cite{Michael2016prx}.  There are also \textit{multimode bosonic codes} that are the superposition states with the same excitation number by combining multiple modes \cite{chuang1997pra,mabuchi1996prl}.
Compared to the conventional multiple-qubits QEC codes, the bosonic QEC modes are hardware-efficient as the logical qubit can be built from a single quantum system with limited error channels like photon loss and pure dephasing. 
Various bosonic codes has been realized in the experiments \cite{Mirrahimi2014njp,Hu2019nature,Fluhmann2019nature,Campagne-Ibarcq2020nature,he2023nc,kudra2022prxq} and have won the break-even point with cat codes \cite{Ofek2016nature} and binomial codes \cite{ni2023nature}.

As both information encoding and error correction continuously consume bosonic code states, it is crucial to prepare high-quality non-Gaussian CV states on demand~\cite{Arne2020prx}, which requires some form of nonlinearity in CV modes. In optical or mechanical systems, due to  the weak accessible nonlinearities (relative to the intrinsic losses), the common strategy  is to combine Gaussian operations with non-Gaussian measurements such as photon number subtraction performed by single photon detectors \cite{hu2023prr,xiang2022npj,liu2022prl}. In contrast, superconducting circuit architectures~\cite{schuster2007nature} leverage Josephson junctions (JJs) to provide strong and controllable nonlinearities, enabling the engineering of non-Gaussian bosonic states. The JJ-based transmon acts as an artificial atomic system and the lowest two levels are usually harnessed to be a physical qubit, which is utilized to implement arbitrary nonlinear phase gates by repeating the noncommuting Rabi gates \cite{park2018njp,park2024npj} or the SNAP gates using a weak drive off-resonant drive \cite{heeres2015prl}. 

For engineering bosonic code states with gate operations, how to decompose the desired arbitrary unitary operation is another complex and challenging problem. 
For example, the SNAP gates inherently rely on an incremental approach to implement phase rotations and strong gates with large rotations would require many repetitions limiting resource efficiency \cite{park2024npj}. In practice, the gate sequence and gate parameters to realize a target operation rely on 
numerical optimization techniques \cite{fosel2020arxiv,kudra2022prxq}.
Recent works have proposed to engineer code states with passive control based on Hamiltonian engineering \cite{Puri2019PRX,Rymarz2021prx,Conrad2021pra,xanda2023arxiv} or reservoir engineering \cite{rojkov2024arxiv}, which can be leveraged to facilitate fault-tolerant operations \cite{Puri2019PRX,Conrad2021pra,xanda2023arxiv,guo2024prl}.

In this work, we propose a new universal gate set that contains only one type of gate operation, which we refer to as the \textit{quantum lattice gate}, cf. Eq.~(\ref{eq-psl}). This gate set is in contrast to the the universal gate set with cubic phase gate, which comprises four distinct gate types with three gate parameters in total, cf. Eq.~(\ref{eq-cubicgate}), and the SNAP gate set, which includes two types of gates but requires an infinite number of gate parameters (depending on the truncated number of Fock states), cf. Eq.~(\ref{eq-SNAP}). Moreover, we provide an analytical deterministic framework to prepare and transform bosonic code states with quantum lattice gates based on Floquet Hamiltonian engineering, and decompose the state preparation/transformation process with sequence of quantum lattices gates.

The structure of the paper is organized as follows. In Sec.~\ref{sec-ugs}, we discuss how to realize universal CV quantum computation and introduce a new universal set of quantum lattice gate that leverages the full nonlinearity of JJ circuits. In Sec.~\ref{sec-bc}, we introduce the definition of fault-tolerant bosonic codes that satisfy the Knill-Laflamme condition, with a focus on cat and binomial code states for the purpose of this work. In Sec.~\ref{sec-gatedecom}, we first classify the code state engineering into three basic processes, i.e., code state preparation, code space embedding, and code space transformation. Then we provide an analytical and deterministic framework to implement these processes, using Hamiltonian engineering combined with an adiabatic ramp protocol. In Sec.~\ref{sec-qlg}, we study how to decompose a given code state engineering process into a sequence of quantum lattice gates. In Sec.~\ref{sec-app}, we apply our method to specific examples, including the preparation of single binomial code states, embedding binomial code space, and transforming binomial codes to cat codes. We also demonstrate automatic error correction for cat states in the presence of single-photon losses. In Sec.~\ref{sec-exp}, we discuss the implementation of quantum lattice gates with a superconducting circuit architecture, achieved by coupling a superconducting cavity to a SQUID device with a tunable effective mutual inductance via a coupler. Finally, we summarize our results in this work and outline potential directions for future research in Sec.~\ref{sec-sum}.

\section{Universal gate sets for CV modes}\label{sec-ugs}

\subsection{Cubic phase gate and SNAP gate}

To achieve universal CV quantum computation, it is essential to implement any unitary transformation between two bosoinc states by using a set of universal quantum gates generated by Hamiltonians that are arbitrary polynomials of the CV mode quadratures $\hat{x}$ and $\hat{p}$~\cite{lloyd1999prl,sefi2011prl}. 
For example, the universal gate set for a single CV mode can be chosen as follows \cite{lloyd1999prl,hillmann2020prl,niklas2022prr}
\bea\label{eq-cubicgate}
\Big\{e^{i\frac{\pi}{2}\hat{a}^\dagger\hat{a}}, e^{it\hat{p}}, e^{is\hat{x}^2},e^{i\gamma \hat{x}^3}\Big\},
\eea
where $\hat{a}\equiv(\hat{x}+i\hat{p})/\sqrt{2\lambda}$ is  the ladder operator with the dimensionless Planck constant $\lambda$ given via $[\hat{x},\hat{p}]=i\lambda$.
This universal gate set includes three Gaussian gates, i.e., the phase rotation gate $e^{i\frac{\pi}{2}\hat{a}^\dagger\hat{a}}$, the displacement gate $e^{it\hat{p}}$, and the squeezing gate $e^{is\hat{x}^2}$, allowing to perform any linear transformations between the CV modes generated by an arbitrary quadratic Hamiltonian. The cubic phase gate $e^{i\gamma \hat{x}^3}$ is a non-Gaussian gate operation that realizes arbitrary nonlinear transformations between the CV modes~\cite{park2018njp,hillmann2020prl,niklas2022prr}.
By using the following commutator-based Lloyd-Braunstein decomposition 
\bea\label{eq-AB}
e^{-[\hat{A},\hat{B}]\delta t}=e^{i\hat{A}\sqrt{\delta t}}e^{i\hat{B}\sqrt{\delta t}}e^{-i\hat{A}\sqrt{\delta t}}e^{-i\hat{B}\sqrt{\delta t}}+O(\delta t^3),
\eea
any desired Hamiltonian term as an arbitrary polynomial of the mode quadratures can be generated~\cite{lloyd1995prl,lloyd1999prl,hillmann2020prl}.
Another commonly used single-mode universal gate set contains the following two types of gates~\cite{Krastanov2015pra} 
\bea\label{eq-SNAP}
\Big\{ e^{it\hat{p}}, \ S[\vec{\theta}]\equiv\sum_{n=0}^{\infty}e^{i\theta_n}|n\rangle\langle n |\Big\},
\eea
where $S[\vec{\theta}]$ is the SNAP gate that endows arbitrary phase $\vec{\theta}=\{\theta_n\}_{n=0}^{\infty}$ to the Fock basis $\{|n\rangle\}_{n=0}^{\infty}$. 

For multiple CV modes, combining single-mode universal gate operations with simple two-mode linear operations, such as the two-mode CSUM gate ($e^{-i\hat{x}_1\hat{p}_2}$) or the CZ gate ($e^{i\hat{x}_1\hat{x}_2}$), is sufficient to achieve universal quantum computation~\cite{lloyd1999prl,niklas2022prr}. This is in stark contrast to performing universal quantum computation with DV qubits. In fact, the set of single-mode Gaussian gates together with CSUM gate or CZ gate compose the Clifford set that allows to perform Clifford quantum computations, which can be efficiently simulated by a classical computer~\cite{gottesman1998heisenberg,bartlett2002prl}. The non-Gaussian gates for single CV modes, e.g., the cubic phase gate and SNAP gate, plays the role of a non-Clifford gate that is crucial to accomplish universal gates with quantum speed-up or quantum advantage.

\begin{figure}[ptb]
\centering
\includegraphics[width=\linewidth]{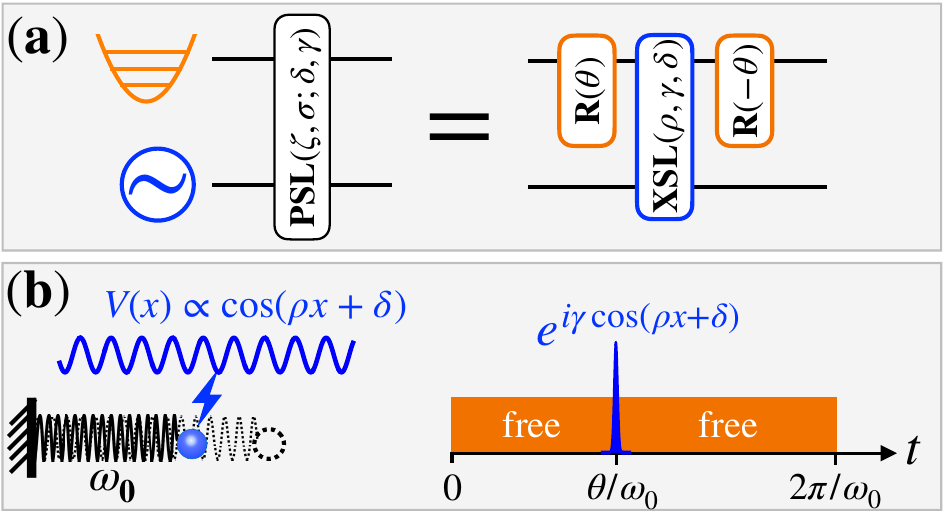}
\caption{Quantum lattice gate. (a) Quantum circuit of a phase-space lattice (PSL) gate decomposed into one x-space lattice (XSL) gate and two phase rotation gates $\mathrm{R(\pm\theta)}$, cf. Eq.~(\ref{eq-PXSL}). The upper and lower lines of circuit represent the bosonic state stored in the cavity and the external driving field exerted on the cavity, respectively.  
(b) Schematic illustration of implementing the PSL gate with a harmonic oscillator with bare frequency $\omega_0$ kicked by a cosine-lattice potential (left). The kick exerted
 at time moment $t=\theta/\omega_0$ realizes the XSL gate $e^{i\gamma \cos(\rho x+\delta)}$; the free evolutions of oscillator before and after the kick realize the phase rotation gates $\mathrm{R(\theta)}$ and $\mathrm{R(-\theta)}$, respectively (right).}
\label{Fig-PXSL}
\end{figure}

\subsection{Quantum lattice gates}

The universal gate set given by Eq.~(\ref{eq-cubicgate}) is comprised of four gate elements with three gate parameters. 
The universal set given by Eq.~(\ref{eq-SNAP}) includes one displacement gate and one SNAP gate with the number of gate parameters  determined by the truncated photon number increasing with the size of the target bosonic state.
In this work, we propose a new universal gate set that only contains one type of elementary gate, i.e.,
 \bea\label{eq-psl}
\Big\{ \mathrm{PSL(\zeta,\sigma; \gamma,\delta)}\equiv e^{i\gamma\cos(\zeta\hat{x}+\sigma\hat{p}+\delta)}\Big\}.
\eea
We refer to this gate as the \textit{phase-space lattice} (PSL) gate, as the generator of such a gate operation is a consine-lattice function in phase space of the CV mode.
We further define an \textit{X-space lattice} (XSL) gate by
\bea\label{eq-XSL}
\mathrm{XSL(\rho,\gamma,\delta)}\equiv e^{i\gamma\cos(\rho\hat{x}+\delta)},
\eea
which is a special case of  $\mathrm{PSL(\zeta,\sigma; \gamma,\delta)}$ gate with gate parameter $\sigma=0$. In view of this, both the PSL and the XSL gates can be referred to as \textit{quantum lattice gates}. 
With the phase rotation gate  $\mathrm{R}(\theta)\equiv e^{-i\theta\hat{a}^\dagger \hat{a}}$,  we can decompose the PSL gate as
 \bea\label{eq-PXSL}
 \mathrm{PSL(\zeta,\sigma; \gamma,\delta)}=\mathrm{R(-\theta)}\mathrm{XSL}(\rho,\gamma,\delta)\mathrm{R(\theta)},
\eea
where the gate parameters $\rho$ and $\theta$ are determined by $\zeta=\rho\cos\theta$ and $\sigma=\rho\sin\theta$. 
We illustrate this gate decomposition with quantum circuit representation in Fig.~\figpanel{Fig-PXSL}{a}. 

To elucidate how the PSL gate can be implemented, we consider a harmonic oscillator with a bare frequency $\omega_0$, as shown in Fig.~\figpanel{Fig-PXSL}{b}. The phase rotation gate $\mathrm{R}(\theta)$ can be simply generated by the free-time evolution with $\theta=\omega_0 t$. Subsequently, the oscillator is kicked by a lattice potential $V(x)\propto\cos(\rho\hat{x}+\delta)$, which results in the XSL gate operation described by Eq.~(\ref{eq-XSL}). Finally, a following free-time evolution over one harmonic period ($\theta=2\pi-\omega_0 t$) completes the process by implementing a phase rotation gate $\mathrm{R}(-\theta)$.
In Section~\ref{sec-exp}, we will discuss further the experimental implementation of quantum lattice gates introduced above with JJ-based superconducting circuit architectures.
For a typical superconducting microwave cavity operating at $\SI{}{\giga\hertz}$ frequencies, our proposed quantum lattice gate operation can be implemented in less than one nanosecond ($<\SI{}{\nano\second}$). In contrast, the typical cubic phase gate requires tens of nanoseconds \cite{eriksson2024nc} while the typical SNAP gate costs several microseconds ($>\SI{}{\micro\second}$) \cite{heeres2015prl,kudra2022prxq} due to the weakness of the dispersive interaction.

\subsection{Nonlinearity as a quantum resource}

The nonlinearity of JJ-based superconducting circuits is a valuable resource for CV quantum information processing and universal quantum computing \cite{hua2024arxiv}. 
To generate a nonlinear cubic phase gate, three-wave mixing is usually employed to pick up the third-order nonlinearity and eliminate the higher-order contributions~\cite{eriksson2024nc}. For the SNAP gate, only the two lowest levels of the transmon are used to interact with the cavity and all other levels are neglected~\cite{heeres2015prl}.  Both the cubic phase and the SNAP gates ignore the higher-order nonlinear terms of the JJ potential~\cite{heeres2015prl,sanchar2022jpm}, and thus do not utilize the full nonlinearity as a quantum resource. Even worse, the residual nonlinearity causes errors as a destructive channel and could become a dominant source of error for engineering large states~\cite{eriksson2024nc}. In contrast, our proposed lattice gates directly utilize the entire nonlinear cosine potential offered by a JJ by coherently harnessing all the high-order nonlinear terms.

\begin{figure}[ptb]
\centering
\includegraphics[width=\linewidth]{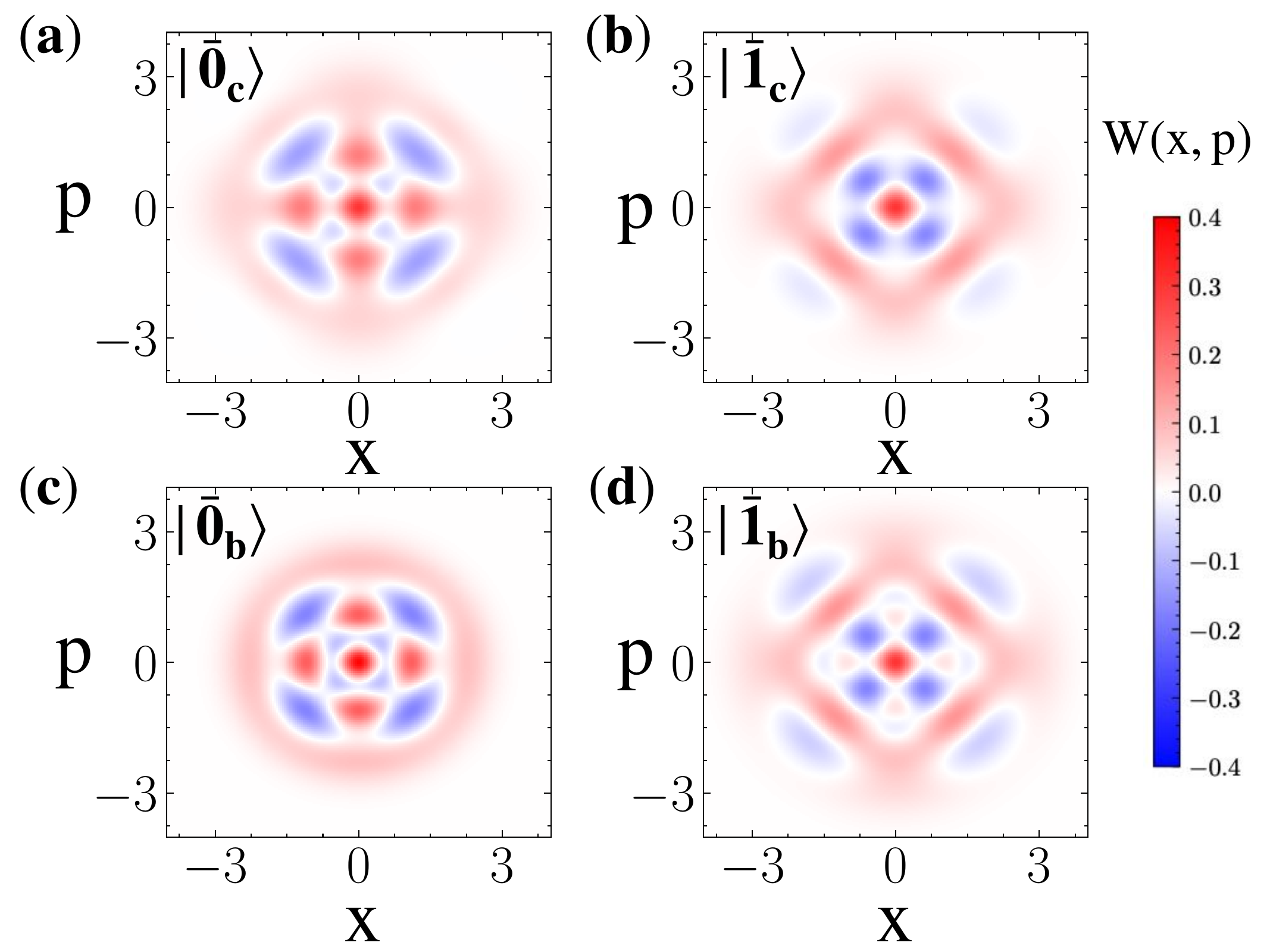}
\caption{Wigner functions of bosonic code states $W(x,p)$. (a)-(b) Two logical cat code states $|\bar{0}_{c}\rangle$ and $|\bar{1}_{c}\rangle$ defined in Eq.~(\ref{eq-sm-4cat})  for the first ``sweet spot'' ($\alpha^2\approx 2.34$) determined by Eq.~(\ref{eq-sm-kl}). (c)-(d) Two logical binomial code states $|\bar{0}_{b}\rangle$ and $|\bar{1}_{b}\rangle$ defined in Eq.~(\ref{eq-bin}).}
\label{Fig-WignerStates}
\end{figure}

\section{Fault-tolerant Bosonic codes}\label{sec-bc}

While the above introduced universal gates allow for arbitrary transformations over CV modes, there is no guarantee of fault tolerance against errors. Due to the intrinsic continuous error channels, e.g., small diffusion along the two quadratures or weak shifts in phase rotations, precise control of CV states is typically more difficult than that of DV states. 
In fact, the no-go theorem claims that the most common and practically relevant Gaussian CV errors cannot be suppressed solely with Gaussian states and Gaussian operations~\cite{niset2009prl}.
It is thus very important to correct errors for practical quantum computations with CV states. 

Similar to QEC using DV systems, fault tolerance always requires some forms of discretization over CVs. 
Bosonic codes provide a solution to this problem by embedding a finite-dimensional code space into the infinite-dimensional Hilbert space of a CV system~\cite{Travaglione2002pra}. 
Compared to QEC codes based on DV systems, bosonic codes are more hardware-efficient, as they leverage the infinite-dimensional Hilbert space of a single CV mode to achieve the necessary redundancy~\cite{weizhou2021fr}.
Bosonic codes also circumvent the no-go theorem of Gaussian CVs, enabling one to detect and correct small errors~\cite{niset2009prl}.
Unlike CV universal quantum computation without error correction, which is similar to noisy intermediate-scale quantum (NISQ) simulators in the context of qubit-based quantum computation~\cite{preskill2018quatum}, bosonic codes enable fault-tolerant digital computation of quantum systems~\cite{sawaya2020npjq}.

The essence of bosonic QEC codes lies in encoding and protecting quantum information within a code space spanned by discretized non-Gaussian states (NGSs), of which two are the code states $| \bar{0}\rangle$ and $| \bar{1}\rangle$. However, not all the NGSs are suitable for bosonic code states for fault-tolerant quantum processing. Precisely speaking, the chosen bosonic code states have to satisfy the so-called Knill-Laflamme condition \cite{knill1997pra,chuang2010book}
\begin{eqnarray}\label{eq-KL}
 \left \{ \begin{array}{lll}
\langle \bar{0}|\hat{\varepsilon}_{i}^{\dag}\hat{\varepsilon}_{j}|\bar{0}\rangle&=&\langle \bar{1}|\hat{\varepsilon}_{i}^{\dag}\hat{\varepsilon}_{j}|\bar{1}\rangle,\\
 \langle \bar{0}|\hat{\varepsilon}_{i}^{\dag}\hat{\varepsilon}_{j}|\bar{1}\rangle&=&\langle \bar{1}|\hat{\varepsilon}_{i}^{\dag}\hat{\varepsilon}_{j}|\bar{0}\rangle=0
 \end{array} \right.
\end{eqnarray}
with the error set $\varepsilon=\{\hat{I}, \hat{\varepsilon}_{1}, \hat{\varepsilon}_{2}, \hat{\varepsilon}_{3},\cdots\}$. For cavity modes, the usual dominant error channels are the single-photon loss $ \hat{\varepsilon}_{1}=\hat{a}$ and dephasing $ \hat{\varepsilon}_{2}=\hat{a}^\dagger\hat{a}$. The Knill-Laflamme condition guarantees that any encoded quantum state  ($|\psi\rangle =c_0|\bar{0}\rangle+c_1|\bar{1}\rangle$) corrupted by error channels  ($|\psi\rangle\rightarrow c_0\hat{\varepsilon}_{i}|\bar{0}\rangle+c_1\hat{\varepsilon}_{j}|\bar{1}\rangle$) is only transformed into another orthogonal basis without losing the quantum information encoded in the superposition coefficients.

For example, the four-legged cat code states~\cite{Mirrahimi2014njp,Leghtas2013prl,Terhal2020iop}
\bea\label{eq-sm-4cat}
 \left \{ \begin{array}{lll}
|\bar{0}_c\rangle\equiv\frac{1}{\sqrt{\mathcal{N}_0}}\big(|\alpha\rangle+|-\alpha\rangle+|i\alpha\rangle+|-i\alpha\rangle\big)\label{eq-sm-4cat1}\\
|\bar{1}_c\rangle\equiv\frac{1}{\sqrt{\mathcal{N}_2}}\big(|\alpha\rangle+|-\alpha\rangle-|i\alpha\rangle-|-i\alpha\rangle\big)\label{eq-sm-4cat2},  
\end{array} \right.
\eea
were proposed to correct single-photon loss with error set $\varepsilon=\{\hat{I}, \hat{a}\}$, where $\mathcal{N}_{m}=8e^{-\alpha^2}[\cosh\alpha^2+(-1)^{\frac{m}{2}}\cos\alpha^2]$ for $ m=0, 2$
is the corresponding normalization factor. However, the four-legged cat code states satisfy the Knill-Laflamme condition only when $\alpha$ is chosen at specific ``sweet spots'' determined by \cite{Terhal2020iop,Joshi2021qst}
\bea\label{eq-sm-kl}
\tan\alpha^2=-\tanh\alpha^2.
\eea
Otherwise, the Knill-Laflamme condition for such cat codes is only approximately satisfied in the limit of large coherent states, i.e., $|\alpha|\gg1$. 

In order to further correct pure dephasing errors described by the error set $\varepsilon=\{\hat{I}, \hat{a}, \hat{n}\}$, where $\hat{n}\equiv \hat{a}^\dagger\hat{a}$ is the photon number operator, one can encode the quantum information into the so-called \textit{binomial codes} \cite{Michael2016prx}
\begin{eqnarray}\label{eq-bin}
 \left \{ \begin{array}{lll}
|\bar{0}_b\rangle&\equiv&\frac{1}{2}\big(|0\rangle+\sqrt{3}|4\rangle\big)\label{eq-bin01}\\
|\bar{1}_b\rangle&\equiv&\frac{1}{2}\big(\sqrt{3}|2\rangle+|6\rangle\big), \label{bistates}
 \end{array}\right.
\end{eqnarray}
where $|n\rangle$ is the Fock state with $n$ photons. 
One can check that the above binomial code states exactly satisfy the Knill-Laflamme condition Eq.~(\ref{eq-KL}) with 
\begin{eqnarray}\label{eq-bin-LF}
 \left \{ \begin{array}{lll}
\langle \bar{0}_b|\hat{n}|\bar{0}_b\rangle=\langle \bar{1}_b|\hat{n}|\bar{1}_b\rangle=3\\
\langle \bar{0}_b|\hat{n}^2|\bar{0}_b\rangle=\langle \bar{1}_b|\hat{n}^2|\bar{1}_b\rangle=12.
 \end{array}\right.
\end{eqnarray}
To correct errors that are polynomials up to a specific degree in bosonic creation and annihilation operators, one has to introduce more complicated binomial code states that are formed by a finite superposition of Fock states weighted with appropriate binomial coefficients~\cite{Michael2016prx}.

In Fig.~\ref{Fig-WignerStates}, we visualize the cat and binomial code states by  plotting their Wigner functions (see the detailed experessions in Appendix~\ref{app-Fourier}). One can find that all the Wigner functions are invariant under a $\frac{\pi}{2}$-rotation in phase space.
In fact, both the cat and binomial code states are the eigenstates of the parity operator 
\bea\label{eq-Parity}
\hat{\cal P}=\text{exp}(i\pi\hat{n})
\eea
with eigenvalue ``+1" (stabilizer). A single-photon loss will change their parity to ``-1", which can be detected by the non-demolition parity measurements~\cite{haroche2007,sun2014nat,rosenblum2018sci} and then can be corrected accordingly. We will discuss the error correction further in Sec.~\ref{sec-aqec}.

\begin{figure}[ptb]
\centering
\includegraphics[width=\linewidth]{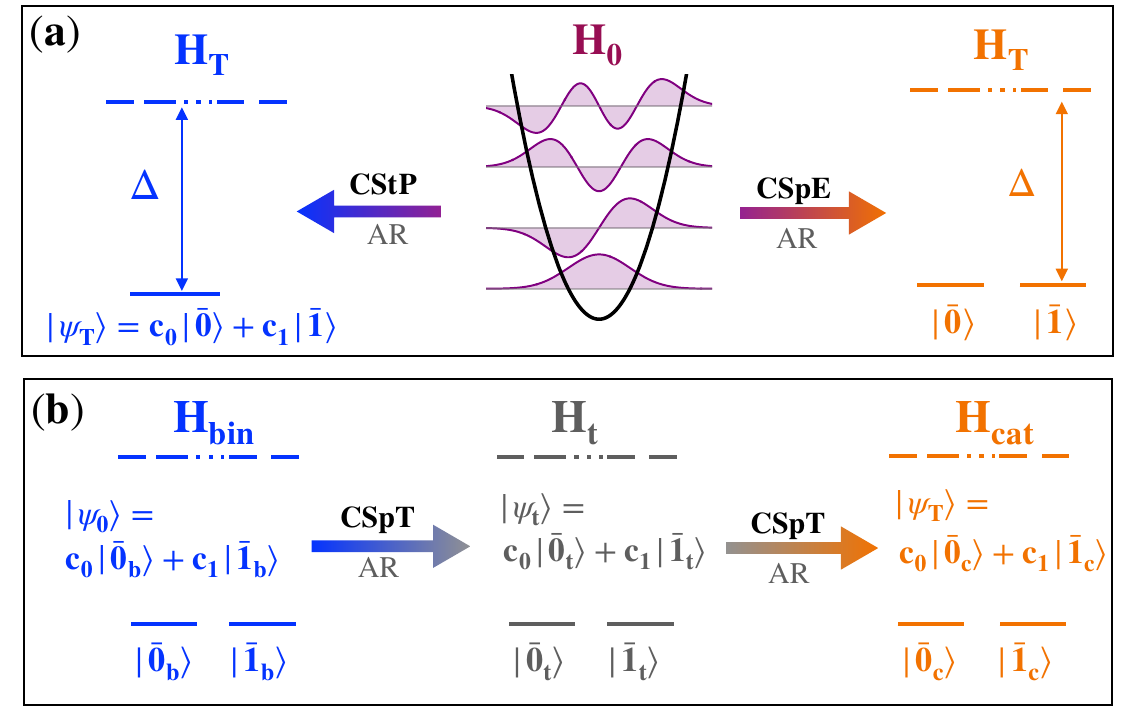}
\caption{Code states engineering.  (a) Code state preparation (CStP) and Code space embedding (CSpE) via adiabatic ramp (AR) from the bare cavity Hamiltonian $\hat{H}_0$ to the target Hamiltonian $\hat{H}_T$ construct by Eq.~(\ref{eq-Hf}) and Eq.~(\ref{eq-HTembed}) respectively. (b) Code space transformation (CSpT) from the binomial Hamiltonian $\hat{H}_{bin}$ to the cat Hamiltonian $\hat{H}_{cat}$ via a transit Hamiltonian $\hat{H}_t$, cf. Eqs.~(\ref{eq-Htran})--(\ref{eq-Hbc}). }
\label{Fig-DesignH}
\end{figure}

\section{Bosonic Code States Engineering }\label{sec-gatedecom}

For fault-tolerant bosonic quantum computation, the ability to engineer bosonic code states is essential. In this section, we study three basic state engineering processes, i.e., \textit{code state preparation}, \textit{code space embedding} and \textit{code space transformation}, as illustrated in Fig.~\ref{Fig-DesignH}.
In most bosonic quantum computing schemes with nonlinear phase gates or SNAP gates, the gate sequences and parameters are determined by various numerical optimization methods~\cite{arrazola2019qst,fosel2020arxiv,landgraf2023fast,zen2024quantum}. 
Here, in contrast, we provide an analytical and deterministic framework for engineering bosonic code states based on Floquet Hamiltonian engineering and the \textit{adiabatic ramp} protocol \cite{xanda2023arxiv}.

\subsection{Code state preparation}
We begin by discussing the process of code state preparation, which typically starts from the ground state of a bare cavity (harmonic oscillator), cf. the middle part in Fig.~\figpanel{Fig-DesignH}{a}, with the bare Hamiltonian 
$
\hat{H}_0=\frac{1}{2}\omega_0(\hat{x}^2+\hat{p}^2).
$
Our first objective is to design a target Hamiltonian $H_T$ whose ``ground state" or eigenstate is the desired target state $|\psi_T\rangle$. For example, $\hat{H}_T=\beta(\cos\hat{x}+\sin\hat{p})$ could be used for the GKP state~\cite{Gottesman2001pra,xanda2023arxiv}, and $\hat{H}_T=\beta (\hat{a}^{\dagger q}-1)(\hat{a}^q-1)$ for the $q$-leg cat state~\cite{Puri2019PRX,guo2024prl}, where $\beta$ represents the amplitude of the designed potential.

However, it is generally not always possible to find a compact Hamiltonian whose eignenstate is an arbitrary target state, such as the binomial code states~\cite{Michael2016prx}. To address this difficulty, we turn to construct the target Hamiltonian directly from the target state by 
\bea\label{eq-Hf}
\hat{H}_T=-\Delta |\psi_T\rangle\langle \psi_T|,
\eea
where $\Delta>0$ is the energy gap of the target Hamiltonian. In this formulation of target Hamiltonian, the target state $|\psi_T\rangle$ has the lowest eigenvalue $-\Delta$, while all other orthogonal states have zero eignevalues.  

To prepare the target code state $|\psi_T\rangle$, we employ the adiabatic ramp protocol~\cite{xanda2023arxiv}, cf. Sec.~\ref{sec-ap}, from the Hamiltonian $\hat{H}_0$ of the bare cavity to the target Hamiltonian $\hat{H}_T$. As a result, the cavity state is adiabatically tuned from the initial vacuum state to the final target state. The parameter $\Delta$ provides a gap protection for the prepared state. This scheme is illustrated by the ``middle-to-left'' process in Fig.~\figpanel{Fig-DesignH}{a}, cf. a concrete example in Sec.~\ref{sec-scsp}.

\subsection{Code space embedding }

One can also embed a finite-dimensional code space, spanned by the two logical code words $|\bar{0}\rangle$ and $|\bar{1}\rangle$, into the infinite-dimensional Fock space of the cavity by constructing the target Hamiltonian as
\bea\label{eq-HTembed}
\hat{H}_T=-\Delta \Big( |\bar{0}\rangle\langle \bar{0}|+ |\bar{1}\rangle\langle \bar{1}|\Big).
\eea
The corresponding Hilbert space forms a degenerate code space spanned by the two logical states, with a nonzero energy gap separating them from the other orthogonal eigenstates.

As discussed above, we adiabatically ramp the system Hamiltonian from $\hat{H}_0$ to $\hat{H}_T$, and prepare the two bosonic code states from the Fock ground state $|0\rangle$ and the first excited state $|1\rangle$, i.e., $|0\rangle\rightarrow |\bar{0}\rangle$ and $|1\rangle\rightarrow |\bar{1}\rangle$. As a result, all the code states can be prepared from a properly superposed initial state of teh Fock ground and first-excited states with a single state engineering protocol. This scheme is illustrated by the ``middle-to-right'' process in Fig.~\figpanel{Fig-DesignH}{a}, cf. a concrete example in Sec.~\ref{sec-embed}.

\subsection{Code space transformation}\label{sec-cspt}

The binomial code states $|\bar{0}_{b}\rangle$ and $|\bar{1}_{b}\rangle$, as defined in Eq.~(\ref{eq-bin}), exactly satisfy the Knill-Laflamme condition for single-photon loss errors, cf. Eq.~(\ref{eq-bin-LF}). In principle, binomial code states corrupted by errors can be perfectly recovered. However, this correction process requires sophisticated unitary operations that facilitate state transfers between the logical code words and the error words~\cite{Michael2016prx}. Implementing such recovery operations remains a challenge with state-of-the-art technological capabilities~\cite{sun2014nat,Michael2016prx}.
In contrast, the cat code words $|\bar{0}_{c}\rangle$ and $|\bar{1}_{c}\rangle$ defined by Eq.~(\ref{eq-sm-4cat}) satisfy the  Knill-Laflamme condition for single-photon loss error on some ``sweet spots'' determined by Eq.~(\ref{eq-sm-kl}). Such cat codes on sweet spots offer the important advantage that the corrupted state will  automatically return back to itself  after four photon losses  (see more details in Appendix~\ref{app-AQEC}). 
By tracking the results of parity measurements and updating our knowledge on the code basis, we can implement the \textit{automatic quantum error correction} (AQEC) protocol against the single-photon loss errors without feedback operations (see detailed discussion in Section ~\ref{sec-aqec}).

Therefore, we introduce the state transformation between the binomial and cat code spaces on demand without deforming the encoded quantum information (preserving the superposition coefficients of the encoded state), i.e.,
\bea
|\psi_b\rangle=c_0|\bar{0}_{b}\rangle+c_1|\bar{1}_{b}\rangle \rightarrow |\psi_c\rangle=c_0|\bar{0}_{c}\rangle+c_1|\bar{1}_{c}\rangle.
\eea
Such a code space transformation process can be achieved by designing the following target transition Hamiltonian
\begin{equation}\label{eq-Htran}
\hat{H}_{T}(t)=-\Delta\Big(|\bar{0}_{t}\rangle\langle\bar{0}_{t}|+|\bar{1}_{t}\rangle\langle\bar{1}_{t}|\Big),
\end{equation}
with time-dependent eigenstates 
\bea\label{eq-transt}
 \left \{ \begin{array}{lll}
|\bar{0}_{t}\rangle&=&\sqrt{1-h(t)}|\bar{0}_{b}\rangle+\sqrt{h(t)}|\bar{0}_{c}\rangle, \label{trans0}\\
|\bar{1}_{t}\rangle&=&\sqrt{1-h(t)}|\bar{1}_{b}\rangle+\sqrt{h(t)}|\bar{1}_{c}\rangle. \label{trans1}
\end{array} \right.
\eea
Here, $h(t)$ is a slow enough time-varying function with $h(t_{0})=0$ and $h(t_{f})=1$ such that 
\bea\label{eq-Hbc}
 \left \{ \begin{array}{lll}
\hat{H}_{\text{bin}}\equiv\hat{H}_T(t_0)=-\Delta\Big(|\bar{0}_{b}\rangle\langle\bar{0}_{b}|+|\bar{1}_{b}\rangle\langle\bar{1}_{b}|\Big)\\
\hat{H}_{\text{cat}}\equiv\hat{H}_T(t_f)=-\Delta\Big(|\bar{0}_{c}\rangle\langle\bar{0}_{c}|+|\bar{1}_{c}\rangle\langle\bar{1}_{c}|\Big). \label{trans1}
\end{array} \right.
\eea
Here, $t_0$ and $t_f$ are the initial and final time moments, respectively. This process allows any encoded state in the binomial code space to be adiabatically transformed to the corresponding encoded state in the cat code space. This scheme is illustrated in Fig.~\figpanel{Fig-DesignH}{b} and the concrete example is given in Sec.~\ref{sec-cst}.

\subsection{Floquet Hamiltonian engineering}

In order to generate the target Hamiltonian $\hat{H}_T$, which is in general an arbitrary function of quadrature operators $\hat{x}$ and $\hat{p}$, we drive the cavity by a periodic external potential $V(\hat{x},t)=V(\hat{x},t+T_d)$ with $T_d=2\pi/\omega_d$, i.e.,
\begin{equation}
\hat{\mathcal{H}}(t)=\frac{\omega_0}{2}\left(\hat{p}^{2}+\hat{x}^{2}\right)+\beta V(\hat{x},t).
\label{rampH}
\end{equation}
A periodically driven system is also called a \textit{Floquet system}~\cite{Floquet1883,Shirley1965pr}. 
By transforming the above Hamiltonian into the rotating frame of frequency $\Omega=2\pi/T$ with $T=nT_d $ $(n\in \mathbb{Z}^+)$, we have 
$\hat{O}(t)\hat{x}\hat{O}^\dagger(t)=\hat{x}\cos (\Omega t)+\hat{p}\sin (\Omega t)$  with time-evolution operator $\hat{O}(t)\equiv e^{i\hat{a}^\dagger\hat{a}\Omega t}$. The transformed Hamiltonian in the rotating frame is given by
\bea\label{eq-Ht}
\hat{H}(t)&\equiv&\hat{O}(t)\hat{\mathcal{H}}(t)\hat{O}^\dagger(t)-i\lambda \hat{O}(t)\dot{\hat{O}}^\dagger(t)\nl
&=& \beta V\Big[\hat{x}\cos (\Omega t)+\hat{p}\sin (\Omega t),t\Big].
\eea
Here,  we have adapted the multi-photon resonance condition $T=2\pi/\omega_0$ or equivalently $\Omega=\omega_0$, i.e., the driving frequency is set to be $n$ times the bare frequency of the harmonic oscillator.

The Flouqet theorem states that the stroboscopic time evolution of a periodic time-varying system is described by a time-independent Floquet Hamiltonian $\hat{H}_F$ determined by  \cite{Floquet1883,Shirley1965pr,Sambe1973pra,Grifoni1998pr,Eckardt2015NJP,Liang2018njp}
\bea\label{eq-HFt0}
\exp\Big(-i\frac{1}{\lambda}\hat{H}_FT\Big)=\mathcal{T}\exp\Big[-i\frac{1}{\lambda}\int_{0}^{T}\hat{H}(t)dt\Big], 
\eea
where $\mathcal{T}$ is the time-ordering operator. 
Under the rotating wave approximation (RWA), the Floquet Hamiltonian $\hat{H}_F$ is just the time-averaged version of $\hat{H}(t)$ over one Floquet period $T$ ~\cite{guo2024prl,Eckardt2015NJP,Mikami2016prb}, i.e.,
\bea\label{eq-h0h1h3}
\lim_{\omega_0/\beta\rightarrow\infty}\hat{H}^{}_F(\hat{x},\hat{p})&=&\frac{1}{T}\int_{0}^{T}dt  \hat{H}(t).\ \ \ 
\eea
By properly engineering the driving potential $V(\hat{x},t)$~\cite{guo2024prl}, the Floquet Hamiltonian $\hat{H}_F(\hat{x},\hat{p})$ can be designed as the target Hamiltonian $\hat{H}_T(\hat{x},\hat{p})$.

For this purpose, we decompose a given target Hamiltonian $\hat{H}_T(\hat{x},\hat{p})$ as a sum of plane-wave operators in the noncommutative phase space~\cite{guo2024prl}, i.e.,
\bea\label{eq-HTxp}
\hat{H}_T(\hat{x},\hat{p})
=
\frac{1}{2\pi}\int \int dk_x dk_pf_T(k_x,k_p)e^{i(k_x\hat{x}+k_p\hat{p})},
\eea
where the \textit{noncommutative Fourier transformation} (NcFT) coefficient in Eq.~(\ref{eq-HTxp}) is given by~\cite{guo2024prl}
\bea\label{eq-fFT}
f_T(k_x,k_p)=\frac{e^{\frac{\lambda}{4}(k^2_x+k^2_p)}}{2\pi}\int\int dxdp H^Q_T(x,p) e^{-i(k_xx+k_pp)}.\nl 
\eea
Here, the integrand $H^Q_T(x,p)=\langle \alpha |\hat{H}_T|\alpha \rangle$ is the Q-function of the target Hamiltonian with $|\alpha\rangle$ the coherent state defined via $\hat{a}|\alpha\rangle=\alpha |\alpha\rangle$, where $\alpha=(x+ip)/\sqrt{2\lambda}$ with $x\equiv\langle \alpha |\hat{x}|\alpha \rangle$ and $p\equiv\langle \alpha |\hat{p}|\alpha \rangle$.

With the NcFT coefficient, one can design the driving potential by superposing a series of cosine-type lattice potentials as~\cite{guo2024prl}
\bea\label{eq-Vxt-2}
V(x,\Omega t)
&=&\int_{-\infty}^{+\infty}A(k, \Omega t)\cos[kx+\phi(k,\Omega t)]dk.\ \ 
\eea
Here, the tunable time-dependent amplitude $A(k,\Omega t)$ and phase $\phi(k,\Omega t)$ are given by
\bea\label{eq-Aphi}
 \left \{ \begin{array}{lll}
 A(k,t)&=&k\Big|f_{T}(k\cos{\Omega t},k\sin{\Omega t})\Big|\\
\phi(k,t)&=&\text{Arg}\Big[f_{T}(k\cos{\Omega t},k\sin{\Omega t})\Big],
\end{array} \right.
\eea
where we have adopted $k_{x}=k\cos{\Omega t}$ and $k_{p}=k\sin{\Omega t}$. 
Each cosine component can be implemented with, e.g., an optical lattice that is formed by laser beams intersecting at an angle in cold-atom experiments \cite{Moritz2003prl,Hadzibabic2004prl,Guo2022prb} or a JJ potential in superconducting circuits \cite{Chen2014prb,Hofheinz2011prl,Chen2011apl}, see more detailed discussions on the experimental implementations in Sec.~\ref{sec-exp}.

\subsection{Adiabatic ramp}\label{sec-ap}

Following the adiabatic ramp protocol~\cite{xanda2023arxiv}, we prepare a target code state according to the Schr\"odinger equation 
\bea\label{eq-pre-psi}
i\lambda\frac{\partial}{\partial t}|\psi_{\text{pre}}(t)\rangle=\hat{H}_{\text{adia}}(t)|\psi_{\text{pre}}(t)\rangle,
\eea
where the Hamiltonian $\hat{H}_{\text{adia}}(t)$ is given by
\begin{equation}
\hat{H}_{\text{adia}}(t)=\frac{\omega_{0}}{2}\left(\hat{p}^{2}+\hat{x}^{2}\right)+\beta(t)V[\hat{x},\Omega(t)t]
\label{rampH}
\end{equation}
with time-dependent amplitude $\beta(t)$ and frequency $\Omega(t)$ for the designed driving potential $V(\hat{x},\Omega t)$, cf. Eq.~(\ref{eq-Vxt-2}). The amplitude $\beta(t)$  [frequency $\Omega(t)$] of the driving potential is tuned from $\beta(t_{0})=0$ [$\Omega(t_{0})\neq\omega_{0}$] to $\beta(t_{f})=\beta_{f}$ [$\Omega(t_{f})=\omega_{0}$] over a preparation time $t_{f}$ with sigmoidal modulations 
\begin{eqnarray}\label{protocol1}
 \left \{  \begin{array}{lll}
\beta(t)&=&\frac{\beta_{f}}{Z_{1}\left[1+e^{-s_{1}(t-t_{c,1})}\right]}-\frac{\beta_{f}}{1+e^{s_{1}t_{c,1}}}, \\
\Omega(t)&=&\Omega(0)+\frac{\omega_{0}-\Omega(0)}{Z_{2}\left[1+e^{-s_{2}(t-t_{c,2})}\right]}-\frac{\omega_{0}}{1+e^{s_{2}t_{c,2}}},\ \ \ 
\end{array} \right .
\end{eqnarray}
where $Z_{j=1,2}=[1+e^{-s_{j}(t_{f}-t_{c,j})}]^{-1}-(1+e^{s_{j}t_{c,j}})^{-1}$
with $t_{c,j}$ and $s_{j}$ being the time centers and slopes of the two adiabatic processes, respectively.

An initial detuning, i.e., $\Omega(t_{0})\neq\omega_{0}$, is set to avoid resonantly driving a linear oscillator at the beginning. This detuning provides a gap in the Floquet spectrum that can suppress the excitation during the adiabatic preparation process~\cite{xanda2023arxiv}. Once the ramped driving field introduces sufficient nonlinearity to the cavity mode, it can be safely tuned to be resonant with the cavity, i.e.,  $\Omega(t_{f})=\omega_{0}$.

\section{Lattice gates decomposition}\label{sec-qlg}

For a given target Hamiltonian,  each cosine component in the engineered driving potential, cf. Eqs.~(\ref{eq-Vxt-2})--(\ref{eq-Aphi}), corresponds to a JJ-based device (e.g., SQUID or SNAIL) in a superconducting circuit architecture~\cite{xanda2023arxiv} or an optical lattice in a cold-atom experimental platform~\cite{Guo2022prb}. 
The engineered driving potential Eq.~(\ref{eq-Vxt-2}) can be approached via a set of discretized cosine-lattice potentials as follows
\bea\label{discrete}
V(x,t)\approx\sum_{n=0}^{N}A(k_n,\Omega t)\Delta k\cos{\left[k_nx+\phi(k_n,\Omega t)\right]},\ \ \ 
\eea
where $k_n=n\Delta k$ is the discretized wavenumber  with $n$ labelling the discretization step.
In principle, one can use $N$ cosine potentials to synthesize the desired driving potential~\cite{guo2024prl}. However, it is challengeable to precisely manipulate multiple cosine-lattice potentials simultaneously in experiments. 
Here, we propose to engineer code states with a single cosine-lattice potential that implements \emph{lattice gate operations}, cf. Eq.~(\ref{eq-psl}). In this section, we provide an analytical framework to decompose the code state engineering process into a sequence of quantum lattice gates.

\begin{figure}
\centering
\includegraphics[width=\linewidth]{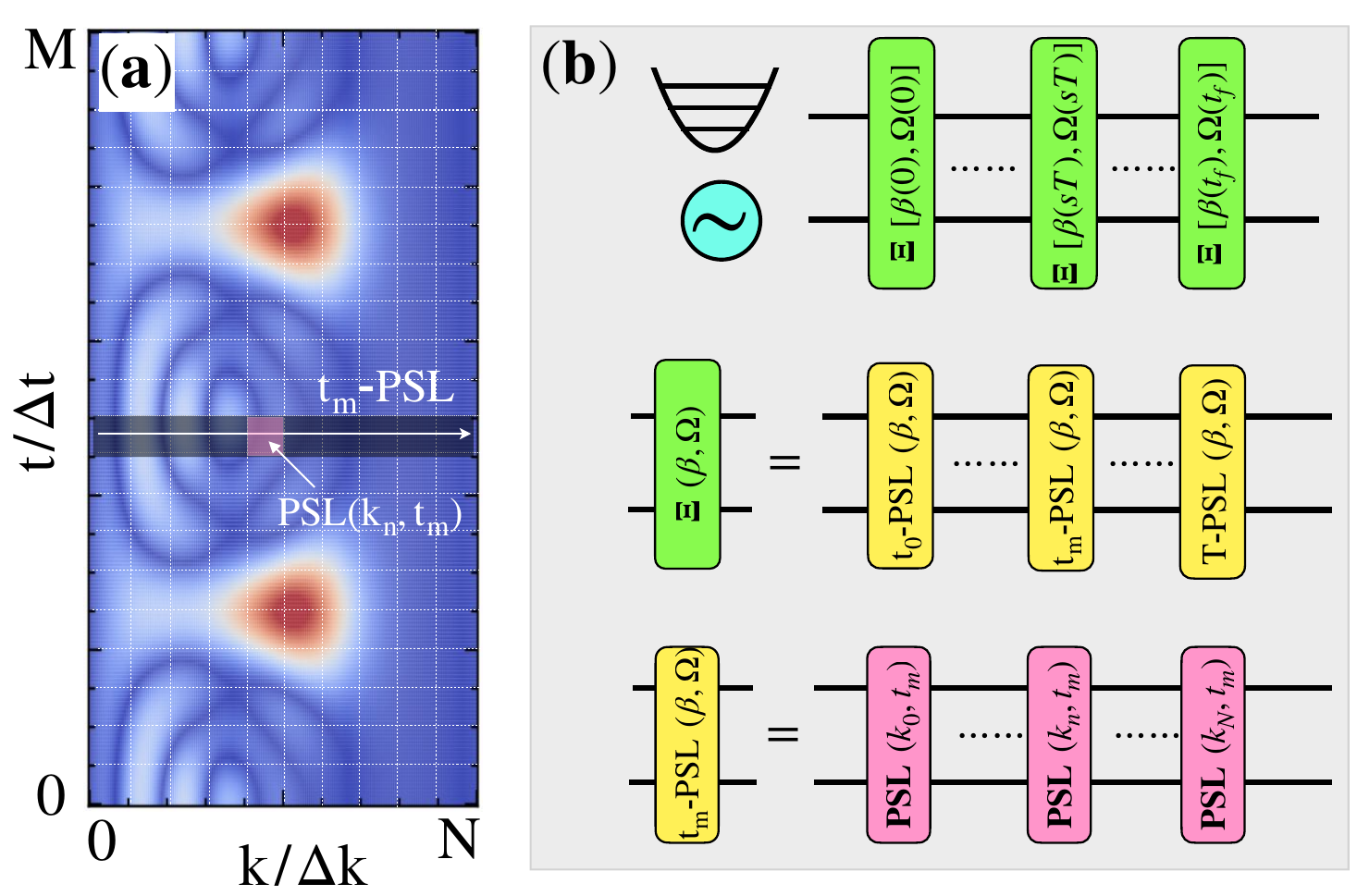}
\caption{Quantum lattice gate decomposition. (a) Chart of driving field modulation, representing the amplitude $A(k_n,t)$ or the phase $\phi(k_n,t)$, over a single Floquet period $T=2\pi/\omega_0$. The whole chart is divided into $N\times M$ grids. Each grid is labelled by $\{k_n=n\Delta k, t_m=m\Delta t\}$ corresponding to a discrete unitary time evolution $\mathrm{PSL}(k_n,t_m)$ that is defined as the \textit{grid lattice gate}, cf. Eqs.~(\ref{eq-Ukntm})--(\ref{eq-pslnm}). The $\mathrm{t_m\mbox{-}PSL}$ gate is composed by concatenating grid lattice gates along the row for a fixed time $t_m$, cf. Eq.~(\ref{eq-PLLtm}). (b) Quantum circuits of lattice gate decomposition: (upper) the whole code state engineering process is decomposed into a sequence of $\mathrm{\Xi}(\beta,\Omega)$ lattice gates, cf. Eq.~(\ref{eq-FloqsT}); (middle) each $\mathrm{\Xi}(\beta,\Omega)$ gate is decomposed into a sequence of $\mathrm{t_m\mbox{-}PSL}$ gates, cf. Eq.~(\ref{eq-PLLtm}); (lower) a single $\mathrm{t_m\mbox{-}PSL}$ gate is decomposed into a sequence of grid lattice gates $\mathrm{PSL}(k_n,t_m)$ shown in (a).}
\label{Fig-GateDecomposition}
\end{figure}

\subsection{Grid lattice gate}

For a given code state engineering process, the driving amplitude $A(k_n,t)$ and phase $\phi(k_n,t)$ in Eq.~(\ref{discrete}) can be calculated according to Eqs.~(\ref{eq-fFT}) and (\ref{eq-Aphi}), as illustrated by a chart in Fig.~\figpanel{Fig-GateDecomposition}{a} over a single Floquet period $T=2\pi/\omega_0$. The chart is divided into small grids with the height and width given by the wavenumber interval $\Delta k=k_{max}/N$ and time step $\Delta t=T/M$ ($N, M\in \mathbb{Z}^+$), respectively. Each grid in the chart, labelled by $\{k_n=n\Delta k, t_m=m\Delta t\}$, corresponds to a discrete unitary time evolution in the rotating frame, cf. Eq.~(\ref{eq-Ht}), i.e.,  
\bea\label{eq-Ukntm}
|\psi(k_n,t_m+\Delta t)\rangle&=&\mathrm{PSL}(k_n,t_m)|\psi(k_n,t_m)\rangle,
\eea
where $\mathrm{PSL}(k_n,t_m)\equiv\mathrm{PSL}(\zeta_{nm},\sigma_{nm}; \gamma_{nm},\delta_{nm})$ is the lattice gate defined in Eq.~(\ref{eq-psl}) with gate parameters
\bea\label{eq-pslnm}
 \left \{ \begin{array}{lll}
\zeta_{nm}=k_n\cos \Omega t_m,&& \sigma_{nm}=k_n\sin \Omega t_m,\ \ \ \\
\gamma_{nm}=-\frac{1}{\lambda}\beta A(k_n,\Omega t_m)\Delta k \Delta t, && \delta_{nm}=\phi\left( k_n,\Omega t_m\right).\ \ \ \ 
\end{array} \right.
\eea
We point out that each discrete unitary time evolution $\mathrm{PSL}(k_n,t_m)$ can be viewed as the operation of a \textit{grid lattice gate}, where the original time parameter just plays the role of a control parameter in this scheme. All gate parameters can be independently tuned based on the restriction condition given by Eq.~(\ref{eq-pslnm}).

\subsection{Gate sequence}

By concatenating the grid lattice gates with discretized wavenumbers for a fixed time step $t_m=m\Delta t$, as shown in Fig.~\ref{Fig-GateDecomposition}(a), we define the lattice gate
\bea\label{eq-PLLtm}
&&\mathrm{t_m\mbox{-}PSL}\equiv\prod_{n=0}^{N}\mathrm{PSL}(k_n,t_m)\\
&&=\prod_{n=0}^{N}e^{-\frac{i}{\lambda}\beta A( k_n,\Omega t_m)\Delta k\Delta t\cos{[ \zeta_{nm}\hat{x}+\sigma_{nm}\hat{p}+\phi(k_n,\Omega t_m)]}}\nl
&&\approx e^{-\frac{i}{\lambda}\Delta t\sum_{n=0}^{N}\beta A(k_n,\Omega t_m)\Delta k\cos{[\zeta_{nm}\hat{x}+\sigma_{nm}\hat{p}+\phi(k_n,\Omega t_m)]}}.\nn
\eea
In the last step, we have adapted the Lie-Trotter product formula~\cite{trotter1959OnTP}, i.e., $e^{i(A+B)}=\lim_{n\rightarrow\infty}(e^{i\frac{1}{n}A}e^{i\frac{1}{n}B})^n$, for sufficiently small time intervals.
Compared to Eq.~(\ref{discrete}) in the rotating frame, cf. Eq.~(\ref{eq-Ht}), the $\mathrm{t_m\mbox{-}PSL}(\beta,\Omega)$ lattice gate faithfully realizes the time evolution with the desired driving potential at each time interval with a single cosine-potential device.

By repeatedly applying the $\mathrm{t_m}$-PSL gate operations at different time steps and concatenating them, we realize the following total gate operation over one full Floquet period
\bea\label{eq-FG}
\mathrm{\Xi}(\beta,\Omega)\equiv\prod_{m=0}^{M}\mathrm{t_m\mbox{-}PSL}(\beta,\Omega),
\eea
which is referred to as the \textit{$\Xi$-type lattice gate} in this paper.
Here, we have explicitly written the Floquet gate with control parameters $\beta$ and $\Omega$. %
In this way, one can prepare the target state by concatenating $\mathrm{\Xi}(\beta,\Omega)$ gates 
\bea\label{eq-FloqsT}
|\psi_f\rangle=\prod_{s=0}^{t_f/T}\mathrm{\Xi}[\beta(sT),\Omega(sT)]|\psi_i\rangle,
\eea
where the adiabatic parameters $\beta(sT)$ and $\Omega(sT)$ with the stroboscopic step index $s=0,1,\cdots,t_f/T$ are determined from the adiabatic ramp protocol, cf. Eq.~(\ref{rampH}). 

In the upper panel of Fig.~\figpanel{Fig-GateDecomposition}{b}, we show the quantum circuits of engineering a sequence of $\mathrm{\Xi}$-type lattice gates for a cavity. The middle panel shows that each  $\mathrm{\Xi}(\beta,\Omega)$ gate is decomposed into a sequence of $\mathrm{t_m\mbox{-}PSL}(\beta,\Omega)$ lattice gates. In the lower panel, we show the quantum circuit of realizing the $\mathrm{t_m\mbox{-}PSL}(\beta,\Omega)$ gate with a sequence of grid lattice gates $\mathrm{PSL}(k_n,t_m)$.

\begin{figure}[ptb]
\centering
\includegraphics[width=\linewidth]{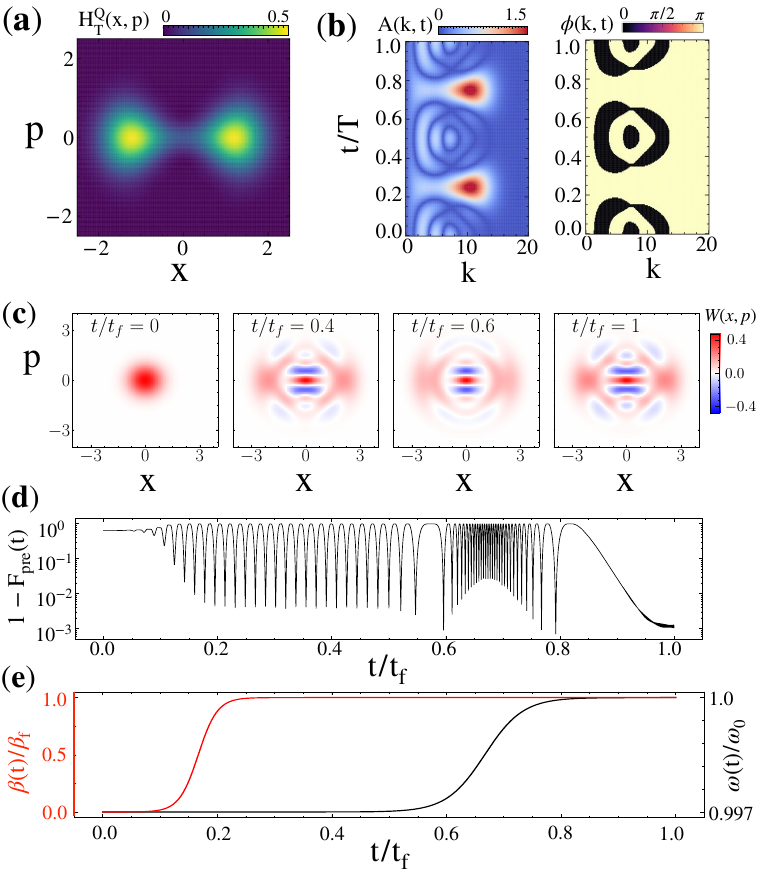}
\caption{State preparation of a single binomial code. (a) Q-function of target Hamiltonian $H^Q_T(x,p)=-\Delta\langle \alpha |\psi_{T}\rangle\langle \psi_{T}|\alpha \rangle$ with $|\psi_{T}\rangle=(|\bar{0}_b\rangle+\sqrt{3}|\bar{1}_b\rangle)/2$.
(b) Charts of driving amplitude $A(k,t)$ (left) and driving phase $\phi(k,t)$ (right) for the engineered driving potential, cf. Eqs.~(\ref{eq-Vxt-2}) and (\ref{eq-Aphi}), for the target Hamiltonian $|\psi_{T}\rangle$. (c) Snapshots of Wigner functions of prepared state $W(x,p)$ during the adiabatic ramp process at different time moments. 
 (d) Time evolution of infidelity of the prepared state with respect to the target state $|\psi_{\text{T}}\rangle$ given by $1-F_{pre}(t)$, where the fidelity $F_{pre}(t)$ is calculated from Eq.~(\ref{eq-fidelity}). 
 (e) Envelopes of driving amplitude $\beta(t)$ and driving frequency $\omega(t)$ during the adiabatic ramp process, cf. Eqs.~(\ref{rampH})--(\ref{protocol1}). Parameters: $\lambda=0.25\omega_{0}$, $\Delta=1.3\omega_{0}$, 
 $\beta_{f}=0.02\omega_{0}$, $\omega(0)=\omega_{0}/(1+\pi\times10^{-3})$, $s_{1}=40/t_{f}$, $s_{2}=30/t_{f}$, $t_{c,1}=t_{f}/6$, $t_{c,2}=2t_{f}/3$.}
\label{Fig-SingleCode}
\end{figure}

\section{Applications}\label{sec-app}

In this section, we apply our method to concrete examples of bosonic code state engineering discussed in Sec.~\ref{sec-gatedecom}. We will first discuss how to prepare a single binomial code state and embed the binomial code space. Then we will show the transformation from binomial codes to cat codes. Finally, we discuss the automatic quantum error correction of the obtained cat code states against single-photon losses.

\begin{figure}[ptb]
\centering
\includegraphics[width=\linewidth]{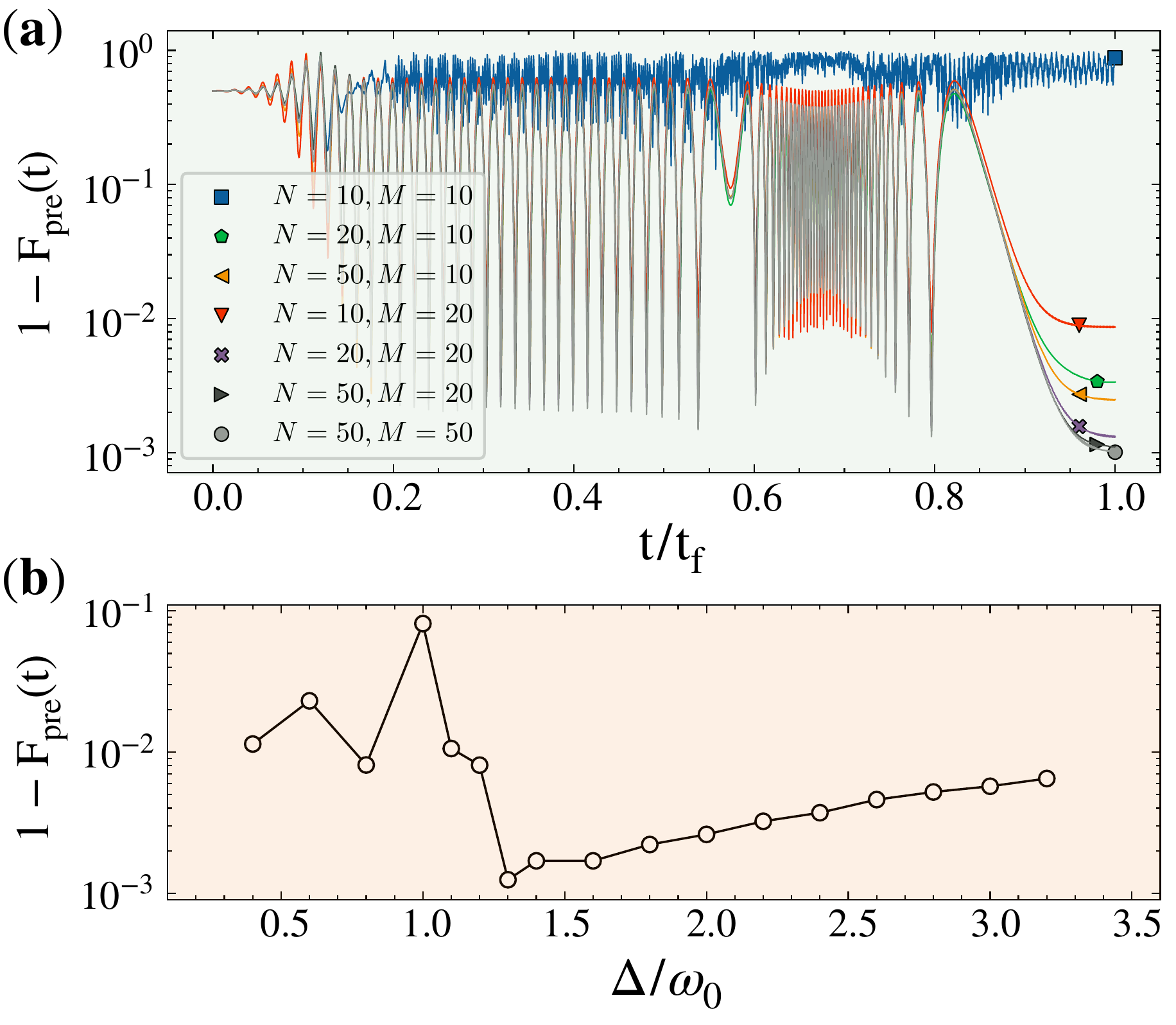}
\caption{ Unitary errors. (a) Infidelity of prepared state $1-F_{pre}(t)$, where the fidelity $F_{pre}(t)$ is calculated from Eq.~(\ref{eq-fidelity}), as a function of time for different wavenumber steps $N$ and time steps $M$. (b) Final infidelity of prepared state, i.e., $1-F_{pre}(t_f)$,  as a function of energy gap $\Delta$ for $N,M=20$. Other system parameters are set to be the same as that in Fig.~(\ref{Fig-SingleCode}).}
\label{Fig-discretization}
\end{figure}

\subsection{Single code state preparation}\label{sec-scsp}
As the first application of our method, we aim to prepare an arbitrary superposition state of binomial code words given by
\begin{equation}
|\psi_{T}\rangle=c_{0}|\bar{0}_b\rangle+c_{1}|\bar{1}_b\rangle.
\label{superposition}
\end{equation}
According to Eq.~(\ref{eq-Hf}), we construct the target Hamiltonian by $\hat{H}_T=-\Delta|\psi_{T}\rangle\langle \psi_{T} |$. From Eq.~(\ref{eq-fFT}), we calculate the NcFT coefficient of the target Hamiltonian as follows (see detailed derivation in Appendix~\ref{app-Fourier})
\begin{eqnarray}\label{newfT}
&&f_{T}(k,\tau)=\\
&&-\frac{\Delta}{4}\Big[|c_{0}|^{2}f_{00}+3|c_{1}|^{2}f_{22}+3|c_{0}|^{2}f_{44}+|c_{1}|^{2}f_{66}\nl
&&+\sqrt{3}\left(|c_{0}|^{2}f_{04}+|c_{1}|^{2}f_{26}+c_{0}c_{1}^{*}f_{02}+c_{0}c_{1}^{*}f_{46}+\text{c.c.}\right)\nl
&&+\left(3c_{1}c_{0}^{*}f_{24}+c_{0}c_{1}^{*}f_{06}+\text{c.c.}\right)\Big]\nn
\end{eqnarray}
with the explicit expression of the Fourier component $f_{nm}$ given by Eq.~(\ref{fTcomponent}) in Appendix.~\ref{app-Fourier}.

Both the general superposition state $|\psi_{T}\rangle$ and the corresponding target Hamiltonian $\hat{H}_T$ are invariant under the two-fold discrete phase-space rotation given by the parity operator $\hat{\cal P}_{}=\text{exp}(i\pi\hat{n})$~\cite{Guo2013prl,Arne2020prx}, cf. Eq.~(\ref{eq-Parity}). 
As an example, we choose the superposition coefficients of target state as $c_0=1/2$ and $c_1=\sqrt{3}/2$ in Eq.~(\ref{superposition}).
In Fig.~\ref{Fig-SingleCode}(a), we plot the Q-function of $\hat{H}_T$ in phase space, i.e., 
$
H^Q_T(x,p)=\langle \alpha |\hat{H}_T|\alpha \rangle,
$
which clearly shows a two-fold rotational symmetry in phase space.
The engineered periodically driving potential can be straightforwardly calculated from Eqs.~(\ref{eq-Vxt-2}), (\ref{eq-Aphi}) and (\ref{newfT}).
In Fig.~\ref{Fig-SingleCode}(b), we plot the time variation of the amplitude $A(k,t)$ and phase $\phi(k,t)$ for the engineered driving potential, cf. Eqs.~(\ref{eq-Vxt-2}) and (\ref{eq-Aphi}). 

We then adiabatically ramp the driving potential following Eq.~(\ref{protocol1}) to prepare the target state. 
 In Fig.~\ref{Fig-SingleCode}(c), we plot the snapshots of the Wigner functions of prepared state during the adiabatic ramp process. 
To quantify the performance of state preparation, we define the fidelity~\cite{james2001pra} of  prepared state with respect to the target code state by 
\begin{equation}
F_{\text{pre}}(t)\equiv\text{Tr}\left[\sqrt{\hat{\rho}_{T}^{1/2}\hat{\rho}_{\text{pre}}(t)\hat{\rho}_{T}^{1/2}}\right],
\label{eq-fidelity}
\end{equation}
where $\hat{\rho}_{pre}=|\psi_{pre}\rangle\langle\psi_{pre}|$ is the density operator of the prepared state and $\hat{\rho}_{T}=|\psi_{T}\rangle\langle\psi_T|$ is the density operator of the target code state. This fidelity measures how closely the prepared state approximates the target state. For pure state, the above defined fidelity reduces to $F_{pre}(t)=|\langle \psi_{pre}(t)|\psi_{T}\rangle|$.
 In Fig.~\ref{Fig-SingleCode}(d), we plot the infidelity of the prepared state $1-F_{pre}(t)$ at stroboscopic time steps during the whole preparation process. The early oscillating behavior of infidelity arises from the initially detuned driving potential that rotates the prepared states in the phase space. This oscillation ceases once the driving potential becomes resonant with the cavity mode.

 \begin{figure*}
\centering
\includegraphics[width=\linewidth]{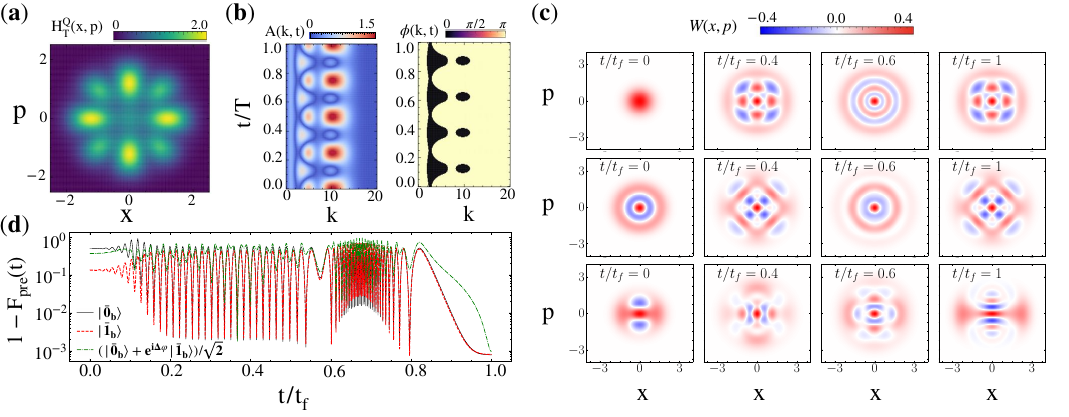}
\caption{Code space embedding. (a) Q-function of target Hamiltonian  given by Eq.~(\ref{eq-HTembedbin}). (b) Charts of driving amplitude $A(k,t)$ and driving phase $\phi(k,t)$ for the target Hamiltonian. (c) Snapshots of Wigner functions of prepared states $W(x,p )$ with target states: (upper) binomial logical state $|\psi_T\rangle=|\bar{0}_b\rangle$, (middle) binomial logical state $|\psi_T\rangle=|\bar{1}_b\rangle$ and (lower) binomial superposition state $|\tilde{\psi}_T\rangle=(|\bar{0}_b\rangle+e^{i\Delta\varphi}|\bar{1}_b\rangle)/\sqrt{2}$, cf. Eq.~(\ref{eq-psiTnew}). (d) Infidelities of three prepared states with respect to the corresponding target states as a function of time during the adiabatic ramp process. For the superposition state, the target state has been updated according to Eq.~(\ref{eq-psiTnew}). Parameters: $\lambda=0.25$, $\Delta=1.4\omega_{0}$, $\beta_{f}=0.02\omega_{0}$, $\omega(0)=\omega_{0}/(1+\pi\times10^{-3})$, $s_{1}=40/t_{f}$, $s_{2}=30/t_{f}$, $t_{c,1}=t_{f}/6$, $t_{c,2}=2t_{f}/3$.
}
\label{Bramp}
\end{figure*}

 \subsubsection{Unitary errors}
 
 In practice, unavoidable unitary errors can occur dur-
ing the state preparation process. Here, we analyse the possible unitary errors in the code state engineering process. The first unitary error source is from the discretization of  implementing the driving potential, cf. in Eq.~(\ref{discrete}), with finite wavenumber interval and time step.  In Fig.~\ref{Fig-discretization}(a), we plot the infidelity of prepared state as a function of time for different number of wavenumber steps $N$ and the number of time steps $M$. As expected, fewer wavenumber and time steps result in larger discretization errors, as shown, e.g., by the result for $N=10$ and $M=10$.  As the number of discretized steps increase, the errors are quickly suppressed and eventually remain almost unchanged for $N,M>20$, see the final infidelity of prepared state for $(N=20,M=20)$ and $(N=50,M=50)$. Hereafter in this paper, unless otherwise specified, we set the wavenumber and time steps to $N,M=20$. 

The second unitary error source is the non-adiabatic excitation that causes leakage of code state to the excited error states during the finite-time adiabatic ramp process. This type of error can, in principle, be suppressed by using a larger energy gap $\Delta$.
To reveal the affect of energy gap $\Delta$, we plot the final infidelity of prepared state as a function of the energy gap in Fig.~\ref{Fig-discretization}(b). Indeed, it shows that the infidelity decreases as the gap increases till an optimal gap value $\Delta \approx 0.5\omega_0$. 
However, the infidelity increases again as the gap continues increasing beyond the optimal value. This is because our method relies on the RWA  (valid for $\beta\ll\omega_{0}$). As the gap of Floquet spectrum increases with the driving strength $\beta$, a large enough energy gap will break the RWA and thus affects the validity of Eq.~(\ref{eq-h0h1h3}).
 The  high-order harmonics of the rotating-frame Hamiltonian $\hat{H}_l(\hat{x},\hat{p})=T^{-1}\int_0^T \hat{H}(t)e^{-il\Omega t}dt$ deviate the engineered Hamiltonian in Floquet-Magnus expansion. To suppress unitary non-RWA errors, one can add additional higher-order driving potentials to approach the target Hamiltonian $\hat{H}_T(\hat{x},\hat{p})$ up to desired precision \cite{xu2024arxiv}.

\subsection{Code space embedding }\label{sec-embed}

As the second application, we aim to embed a finite binomial code space into the infinite Fock space of a cavity. According to Eq.~(\ref{eq-HTembed}), the target Hamiltonian can be set as
\bea\label{eq-HTembedbin}
\hat{H}_T=-\Delta \Big( |\bar{0}_b\rangle\langle \bar{0}_b|+ |\bar{1}_b\rangle\langle \bar{1}_b|\Big).
\eea
In Fig.~\ref{Bramp}(a), we calculate and plot  the Q-function $H^{T}_{Q}$ of target Hamiltonian $
H^Q_T(x,p)=\langle \alpha |\hat{H}_T|\alpha \rangle
$. In contrast to the case of single-code state preparation, the current target Hamiltonian Q-function is invariant under four-fold discrete rotations in phase space~\cite{Guo2013prl,Arne2020prx}.
The NcFT Fourier coefficient is given by (see details in Appendix.~\ref{app-Fourier})
\bea
&&f_{T}(k,\tau)=\\
&&-\frac{\Delta}{4}\left[f_{00}+3f_{22}+3f_{44}+f_{66}+\sqrt{3}\left(f_{04}+f_{26}+\text{c.c.}\right)\right],\nn
\label{coefficient}
\eea
which yields the driving field with the charts of periodically modulated amplitude and phase shown in Fig.~\figpanel{Bramp}{b}.

As claimed above, the code space embedding is supposed to enable preparing any superposition state with the same adiabatically ramped driving protocol. 
In the upper and middle rows of Fig.~\figpanel{Bramp}{c}, we show the Wigner functions of two prepared logical code states $|\bar{0}_b\rangle$ and $|\bar{1}_b\rangle$ at different moments, starting from the cavity vacuum state $|0\rangle$ and the second excited Fock state $|2\rangle$, respectively. 
In Fig.~\figpanel{Bramp}{d}, we also plot the time evolution of infidelities for both binomial code words. It is clear that the prepared state eventually shows a very high fidelity as expected. 
\begin{figure}[ptb]
\centering
\includegraphics[width=\linewidth]{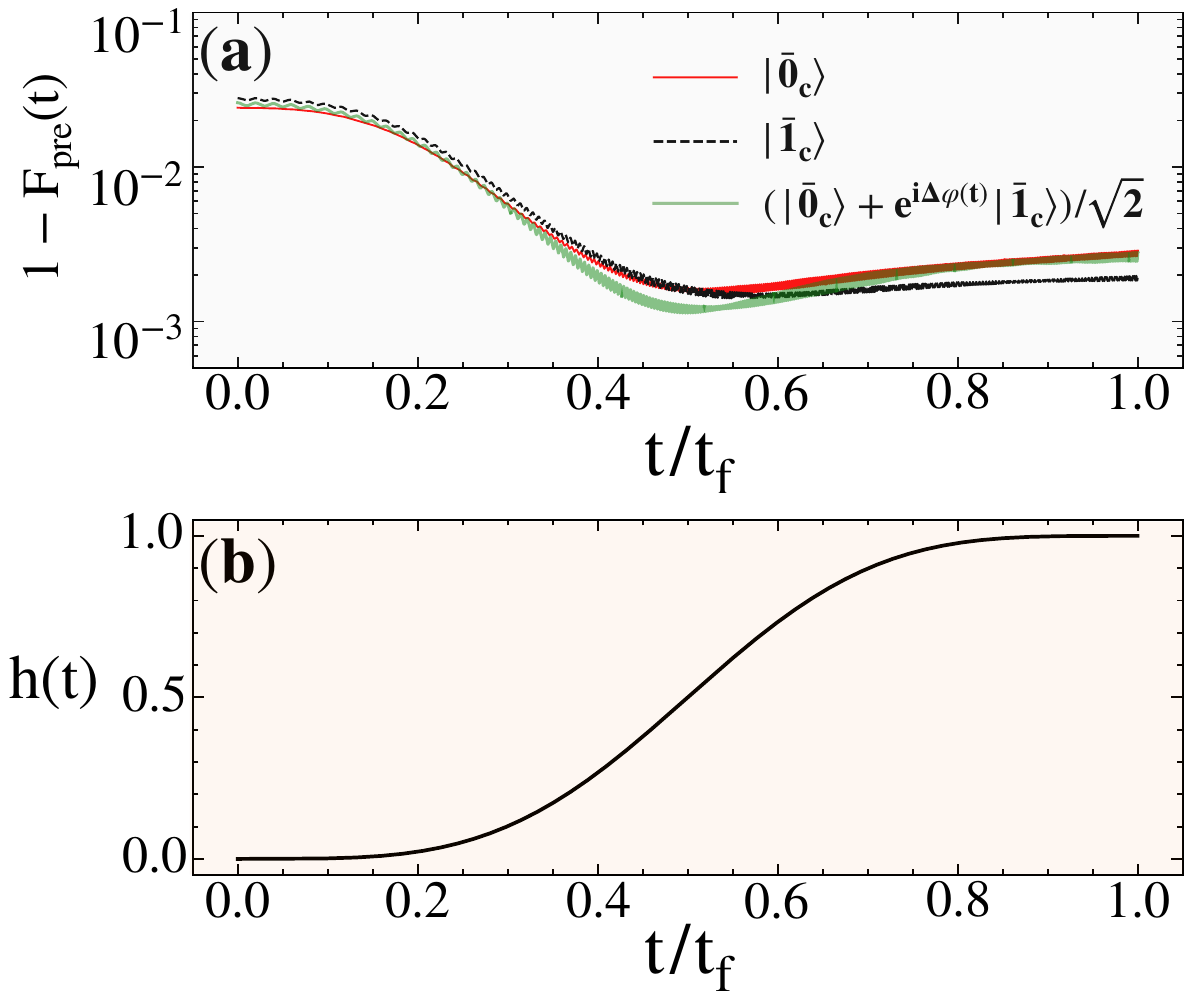}
\caption{Code space transformation. (a) Infidelity time evolution of prepared code states with respect to the target cat code basis $|\psi_T\rangle=|\bar{0}_c\rangle$, $|\psi_T\rangle=|\bar{1}_c\rangle$ and the time-dependent updated superposition state $|\tilde{\psi}_{T}(t)\rangle=( |\bar{0}_c\rangle+e^{i\Delta\varphi(t)}|\bar{1}_c\rangle)/\sqrt{2}$, cf. Eq.~(\ref{eq-psiTnewt}). (b) Envelope of  time-varying function $h(t)$ introduced in the transition Hamiltonian, cf. Eq.~(\ref{eq-transt}). Parameters: $\lambda=0.25$, $t_f = 5000\times 2\pi/\omega_0$, $\Delta = 0.1\omega_0$.  }
\label{BC0}
\end{figure}
In the lower row of Fig.~\figpanel{Bramp}{c}, we show the Wigner functions of a prepared superposition state starting from the superposition Fock state $|\psi_i\rangle=c_0|0\rangle+c_1|2\rangle$ (with superposition coefficients $c_{0}=c_{1}=1/\sqrt{2}$) at different moments. 

It is important to note that the two prepared code words (i.e., $|\bar{0}_{b}\rangle$ and $|\bar{1}_{b}\rangle$) may accumulate different phase factors during the adiabatic preparation process, i.e.,
\bea
|\psi_i\rangle \rightarrow |\psi_{\text{pre}}\rangle=c_{0}e^{i\varphi_0}|\bar{0}_b\rangle+c_{1}e^{i\varphi_1}|\bar{1}_b\rangle.
\eea
Each phase of code words includes both the dynamical and geometrical contributions. Since the embedded code space is degenerate, the phase difference between the two code words $\Delta \varphi=\varphi_1-\varphi_0$ is only determined by the geometric phases, which can be deterministically extracted via $\Delta \varphi=\arg(\langle\bar{1}_b|\psi_{\text{pre}}\rangle \langle\psi_{\text{pre}}|\bar{0}_b\rangle)$ from numerical simulation.
In our case, the extracted geometric phase difference is $\Delta \varphi\approx0.27\pi$, and it remains independent of the superposition coefficients $c_0$ and $c_1$ within our numerical precision. This phase difference also accounts for the slight asymmetry observed in the final Wigner function, cf. the lower row of Fig.~\figpanel{Bramp}{c}. By updating our knowledge of the final target state with the extracted phase difference, i.e., 
\bea\label{eq-psiTnew}
|\tilde{\psi}_{{T}}\rangle\rightarrow c_{0}|\bar{0}_b\rangle+c_{1}e^{i\Delta\varphi}|\bar{1}_b\rangle,
\eea
 the prepared states consistently exhibit high fidelities, as shown in Fig.~\figpanel{Bramp}{d}.

\subsection{Code spaces transformation}\label{sec-cst}

As the final application, we study the code space transformation between the binomial and cat code states by engineering 
the transition Hamiltonian given by Eqs.~(\ref{eq-Htran})--(\ref{eq-transt}).
In Appendix~\ref{app-cst}, we calculate explicitly the Husmi-Q function and the NcFT coefficients for the transition Hamiltonian given by Eq.~(\ref{eq-Htran}).
We consider the slow enough time-varying function $h(t)$ in Eq.~(\ref{eq-transt}) as follows
\bea
h(t)=\sin\left[\frac{\pi}{2}\sin^2\left(\frac{\pi t}{2t_f}\right)\right]^2.
\eea
This function meets the boundary conditions $h(0)=0$ and $h(t_f)=1$
, enabling a state transfer process in adiabatic limit.

In Fig.~\ref{BC0}, we first present numerical simulations of the transformation processes from the binomial code state $|\bar{0}_{b}\rangle$ ($|\bar{1}_{b}\rangle$) to the cat code state $|\bar{0}_{c}\rangle$ ($|\bar{1}_{c}\rangle$). The infidelity behaviors of these processes are respectively depicted by the black and red curves in Fig.~\figpanel{BC0}{a}, while the time evolution of $h(t)$ is shown in Fig.~\figpanel{BC0}{b}. 
Note that the fidelities exhibit relatively high values even at the start of the transformation processes, which can be attributed to the significant overlap between the initial binomial state and the corresponding cat state at the first ``sweet spot'' ($\alpha^2\approx 2.34$) determined by Eq.~(\ref{eq-sm-kl}). This overlap is also reflected in the similarity between the Wigner functions of binomial and cat code states shown in Fig.~\ref{Fig-WignerStates}. Despite this large overlap, the fidelities continue to improve throughout the transformation process, reaching even higher values as they approach their respective target states.

As discussed in Sec.~\ref{sec-embed}, during the adiabatic transition processes, the two logical code words may accumulate different phases. As a result, 
an initial superposition binomial code state $|\psi_{0}\rangle=c_{0}|\bar{0}_{b}\rangle+c_{1}|\bar{1}_{b}\rangle$ is transferred to the following code state
$$|\psi_{t}\rangle= c_{0}e^{i\varphi_0(t)}|\bar{0}_{t}\rangle+c_{1}e^{i\varphi_1(t)}|\bar{1}_{t}\rangle,$$ 
where the transition code words $|\bar{0}_{t}\rangle$ and $|\bar{1}_{t}\rangle$ are given by Eq.~(\ref{eq-transt}). The time-dependent accumulated  phases $\varphi_0(t)$ and  $\varphi_1(t)$ include both dynamical and geometrical contributions. Due to the designed form of transition Hamiltonian given by Eq.~(\ref{eq-Htran}), the transition code words $|\bar{0}_{t}\rangle$ and $|\bar{1}_{t}\rangle$ are degenerate. Thus, the phase difference $\Delta\varphi(t)=\varphi_1(t)-\varphi_0(t)$ 
only contains geometrical contribution and can be deterministically extracted from numerical simulation via $\Delta \varphi(t)=\arg(\langle\bar{1}_t|\psi_{t}\rangle \langle\psi_{t}|\bar{0}_t\rangle)$, which is independent of the superposition coefficients $c_0$ and $c_1$. By updating the target state with the extracted phase difference as a function of time (up to a global phase factor), i.e., 
\bea\label{eq-psiTnewt}
|\tilde{\psi}_{T}(t)\rangle\propto c_{0}|\bar{0}_c\rangle+c_{1}e^{i\Delta\varphi(t)}|\bar{1}_c\rangle,
\eea 
we display  in Fig.~\ref{BC0}(a) the infidelity of the transferred superposition code state  (green curve) exhibiting high fidelities after the transition process.

\subsection{Automatic quantum error correction (AQEC)}\label{sec-aqec}

\begin{figure}[ptb]
\centering
\includegraphics[width=\linewidth]{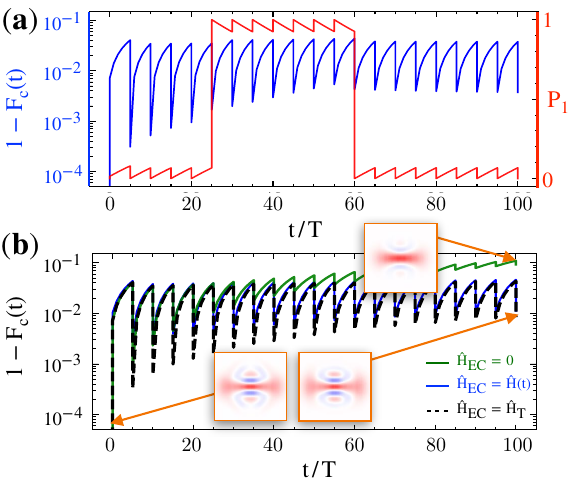}
\caption{ Automatic quantum error correction. (a) The infidelity $1-F_c(t)$ of the protected code state (blue solid) and the parity probability $P_1=\langle\hat{\Pi}_1\rangle$ (red solid) as functions of time for a single quantum trajectory with the driving Hamiltonian protection, i.e., $\hat{H}_{EC}=\hat{H}(t)$ in Eq.~(\ref{eq-EC-2}) with $\beta=0.02\omega_0$.  (b) The infidelity of the protected code state, averaged over $10^3$ quantum trajectories, in the cases of no Hamiltonian protection ($\hat{H}_{EC}=0$; green solid), the ideal Hamiltonian protection [$\hat{H}_{EC}=\hat{H}_T$ in Eq.~(\ref{eq-Htran4}) with $\Delta=0.2\omega_0$; black dashed], and the driving Hamiltonian protection [$\hat{H}_{EC}=\hat{H}(t)$ in Eq.~(\ref{eq-EC-2}) with $\beta=0.02\omega_0$; blue solid].
In both panels, we start from the initial code state $|\psi_c(0)\rangle=\big(\sqrt{5}|\bar{0}_c\rangle+\sqrt{3}|\bar{1}_c\rangle\big)/2\sqrt{2}$ and perform parity measurements every five Floquet periods.
The three insets in (b) show the Wigner functions of code states at the beginning and the end of the time evolution when the updated target states after the last parity measurements recover the initial code state with different AQEC protocols.
Other parameters: $\lambda=0.25$, $\kappa = 10^{-3}\omega_{0}$, wavenumber steps $M=20$ and time steps $N=100$, cf. Fig.~\ref{Fig-GateDecomposition}(a).}
\label{Fig-AutoQEC}
\end{figure}

As mentioned in Section~\ref{sec-cspt}, QEC schemes with binomial codes typically require active error-syndrome measurements and adaptive recovery operations, which are hardware intensive and prone to propagating errors of gate operations. In contrast, cat codes allow for AQEC schemes 
that can protect the encoded quantum information by autonomously correcting photon-loss errors based on parity measurements, without the need for feedback operations \cite{Gertler2021nature}.
In this section, we study in detail how to implement AQEC against single-photon loss errors using four-legged cat states based on our driving protocol.

To investigate the impact of a noisy environment and the error correction process, we extend the unitary Schr\"odinger equation, cf. Eq.~(\ref{eq-pre-psi}), to the Lindblad master equation in the rotating frame as follows
\bea
\frac{\partial}{\partial t}\hat{\rho}_c(t)=&&-\frac{i}{\lambda}\left[\hat{H}_{EC}, \hat{\rho}_{c}(t)\right]+\kappa\mathcal{L}[\hat{a}]\hat{\rho}_{c}(t),\ \ 
\label{Lindblad}
\eea
where $\hat{H}_{EC}$ is the Hamiltonian used for error correction and $\hat{\rho}_{c}(t)=|\psi_c(t)\rangle\langle \psi_c(t)|$ is the density operator of the code state. The second term on the right-hand side of Eq.~(\ref{Lindblad}) describes the decay of the cavity with single-photon loss rate $\kappa$, where $\mathcal{L}[\hat{O}]$ is the Lindblad superoperator defined via $\mathcal{L}[\hat{O}]\hat{\rho}=\hat{O}\hat{\rho}\hat{O}^{\dag}-(\hat{O}^{\dag}\hat{O}\hat{\rho}+\hat{\rho}\hat{O}^{\dag}\hat{O})/2$.

Following our method, to protect the quantum information encoded with  cat states, we set the target Hamiltonian to be
\begin{equation}\label{eq-Htran4}
\hat{H}_{T}=-\Delta\Big(|\bar{0}_{c}\rangle\langle\bar{0}_{c}|+|\bar{1}_{c}\rangle\langle\bar{1}_{c}|+|\bar{0}_{e}\rangle\langle\bar{0}_{e}|+|\bar{1}_{e}\rangle\langle\bar{1}_{e}|\Big).
\end{equation}
Here, $|\bar{0}_{c}\rangle$ and $|\bar{1}_{c}\rangle$ are the cat code words defined in Eq.~(\ref{eq-sm-4cat}), while $|\bar{0}_e\rangle\propto\hat{a}|\bar{0}_c\rangle$ and $|\bar{1}_e\rangle\propto\hat{a}|\bar{1}_c\rangle$ are the cat error states defined as
\bea\label{eq-sm-4cat-erorr}
 \left \{ \begin{array}{lll}
|\bar{0}_e\rangle\equiv\frac{1}{\sqrt{\mathcal{N}_{1}}}\big(|\alpha\rangle-|-\alpha\rangle-i|i\alpha\rangle+i|-i\alpha\rangle\big),\label{eq-sm-4cat1-e}\\
|\bar{1}_e\rangle\equiv\frac{1}{\sqrt{\mathcal{N}_{3}}}\big(|\alpha\rangle-|-\alpha\rangle+i|i\alpha\rangle-i|-i\alpha\rangle\big)\label{eq-sm-4cat2-e}  
\end{array} \right.
\eea
with $\mathcal{N}_{m}=8e^{-\alpha^2}[\sinh\alpha^2+(-1)^{\frac{m-1}{2}}\sin\alpha^2]$ for $m=1, 3$ the normalization factors, similar to $\mathcal{N}_{m}$ for $m=0, 2$ in Eq.~(\ref{eq-sm-4cat}). 
We begin with an unknown state that is an arbitrary superposition of the cat code words
\bea\label{eq-psi-c}
|\psi_c(0)\rangle=c_1|\bar{0}_c\rangle+c_2|\bar{1}_c\rangle,
\eea
which is the target state we aim to protect with AQEC.

The first step of our AQEC protocol is to periodically perform the projective parity measurement~ \cite{haroche2007,sun2014nat,rosenblum2018sci} 
\bea\label{eq-PIm}
\hat{\Pi}_{m}=\sum_{n=0}^{\infty}|2n+m\rangle\langle 2n+m|
\eea
with the parity index $m=0, 1$. After each parity measurement, the code state collapses into a state with a certain parity. The probability of obtaining the $m$-parity outcome is given by
\bea\label{eq-Pmt}
P_{m}(t)=\text{Tr}[\hat{\Pi}_{m}\hat{\rho}_c(t)\hat{\Pi}_{m}^{\dag}] 
\eea
with the constraint $P_0+P_1=1$.
Next, we divide the unit interval $[0,1]$ into two sections with lengths $P_0$ and $P_1$, respectively. Then one can generate a random number $\varepsilon_{r}\in[0,1]$ and identify its position within the unit interval. If the random number falls within the $m$-th section, the code state is projected into
\begin{equation}
\hat{\rho}_c(t)\rightarrow\frac{\hat{\Pi}_{m}\hat{\rho}_c(t)\hat{\Pi}_{m}}{P_{m}(t)},
\label{project1}
\end{equation}
and the fidelity is updated as
\begin{equation}
F_c(t)\rightarrow\text{Tr}\left[\sqrt{\hat{\rho}_{T}^{1/2}\hat{\rho}_{c}(t)\hat{\rho}_{T}^{1/2}}\right].
\label{upF}
\end{equation}
Here, $\hat{\rho}_{T}=|\psi_{T}\rangle\langle\psi_{T}|$ is the target density operator with $|\psi_{T}\rangle=|\psi_c(0)\rangle$, cf. Eq.~(\ref{eq-psi-c}). The target state is updated in sequence as follows
\begin{align}
|\psi_T\rangle&=c_1|\bar{0}_c\rangle+c_2|\bar{1}_c\rangle \nn\\
&\rightarrow c_1|\bar{1}_e\rangle+c_2|\bar{0}_e\rangle \nn\\    
&\rightarrow c_1|\bar{1}_c\rangle+c_2|\bar{0}_c\rangle \label{eq-update}\\      
&\rightarrow c_1|\bar{0}_e\rangle+c_2|\bar{1}_e\rangle 
\ \smash{%
\begin{tikzpicture}
\path (0.0,1.5) coordinate(c1)
(0.75,1.0) coordinate(c2)
(0.75,0.5) coordinate(c3)
(0.0,0.0) coordinate(c4);
\draw [-latex] (c4) .. controls (c3) and (c2) .. (c1);
\end{tikzpicture}%
}\nn 
\end{align}%
whenever a parity change is detected (see Appendix~\ref{app-AQEC} for more details). By repeatedly performing parity measurements and updating the target state accordingly, the AQEC protocol is thus implemented using the cat  states.

According to our proposal, the error correction Hamiltonian $\hat{H}_{EC}$ in Eq.~(\ref{Lindblad}) is set by the following time-dependent Hamiltonian, cf. Eq.~(\ref{eq-Ht}),
\bea\label{eq-EC-2}
\hat{H}_{EC}=\hat{H}(t)
&=& \beta V\Big[\hat{x}\cos (\Omega t)+\hat{p}\sin (\Omega t),t\Big],
\eea 
where the driving field $V(x,t)$ is designed from Eqs.~(\ref{eq-Vxt-2})--(\ref{eq-Aphi}) with the target Hamiltonian given in Eq.~(\ref{eq-Htran4}).
From Eqs.~(\ref{eq-PLLtm})--(\ref{eq-FG}), the unitary time evolution over each Floquet period, i.e., $t\in[sT,(s+1)T]$ with $s\in\mathbb{N}^0$ (nonnegative integers), can be decomposed into a sequence of quantum lattice gate operations as described by Eqs.~(\ref{eq-PLLtm}) and (\ref{eq-FG}
).

In Fig.~\ref{Fig-AutoQEC}, we present the results of AQEC starting from a generic encoded state $|\psi_c(0)\rangle=\frac{1}{2}\Big(\sqrt{\frac{5}{2}}|\bar{0}_c\rangle+\sqrt{\frac{3}{2}}|\bar{1}_c\rangle\Big)$. The parity measurement is performed every five Floquet periods.
In Fig.~\figpanel{Fig-AutoQEC}{a}, we plot the infidelity and parity probability $P_1$ after each measurement for a single quantum trajectory with the Hamiltonian protection. It clearly shows that the infidelity decreases suddenly by nearly two orders of magnitude after each parity measurement, while it increases gradually between two successive measurements due to the photon loss, which causes the quantum state to evolve towards a mixed state. 
In Fig.~\figpanel{Fig-AutoQEC}{b}, we plot the infidelity of the protected state averaged over $10^3$ trajectories. For comparison, we also show the result without the Hamiltonian protection (i.e., $\beta=0$). It is clear that the fidelity of the protected state after each parity measurement is significantly enhanced by the Hamiltonian protection. We also plot the Wigner functions of the initial code state $|\psi_c(0)\rangle$ and the code states at the end of the time evolution when the updated target states after the last parity measurements recover the initial code state  for different AQEC protocols, as shown by the insets of Fig.~\figpanel{Fig-AutoQEC}{b}.

In our driving protocol, the Floquet Hamiltonian $\hat{H}_F$, corresponding to the time-periodic Hamiltonian in Eq.~(\ref{eq-EC-2}), approximates the target Hamiltonian $\hat{H}_T$ given by Eq.~(\ref{eq-Htran4}) under the RWA, cf. Eq.~(\ref{eq-h0h1h3}). To assess the errors induced by the RWA, we also show the results for the ideal AQEC protocol, where the error correction Hamiltonian in the quantum master equation (\ref{Lindblad}) is set exactly as the target Hamiltonian~(\ref{eq-Htran4}), i.e., 
$
\hat{H}_{EC}=\hat{H}_T
$, which establishes a lower boundary for the  infidelity.  Our results show that the AQEC performance with our driving Hamiltonian protection nearly overlaps that with the ideal Hamiltonian protection. The  slowly increasing infidelity after each measurement  arises from the multi-photon loss, which cannot be corrected by the four-fold symmetric cat code states~\cite{Leghtas2013prl}. Correcting multi-photon loss errors would require cat code states with higher-order discrete rotational symmetry in phase space~\cite{Arne2020prx}. Another error source arises from the fact that the cat error states $|\bar{0}_e\rangle$ and $|\bar{1}_e\rangle$ do not satisfy the Knill-Laflamme condition at the first sweet spot of the cat code states. This error can be mitigated by choosing cat codes at higher-order sweet spots (i.e., with larger $\alpha$). However, since the decay rate is proportional to the average photon number, this comes at the cost of increasing the frequency of parity measurements.

\section{Discussions}

\subsection{~Experimental implementation}\label{sec-exp}

In principle, the quantum lattice gate operations introduced in Eqs.~(\ref{eq-psl}) and (\ref{eq-XSL}) can be directly demonstrated with a cold atom in the optical lattice potential. However, for practical scalable quantum computing, we propose to implement the quantum lattice gates in superconducting circuit QED architecture. By rapidly turning on/off the capacitive coupling between the cavity and the SQUID or SNAIL circuit, one can implement the X-space lattice gate  
$
\mathrm{XSL(1,\gamma,\delta)}= e^{i\gamma\cos(\hat{\varphi}+\delta)}.
$
To implement a more general X-space lattice gate $\mathrm{XSL(\rho,\gamma,\delta)}= e^{i\gamma\cos(\rho\hat{\varphi}+\delta)}$, as given by Eq.~(\ref{eq-XSL}), one can apply additional squeezing operations, i.e., 
\bea\label{eq-XSL-1}
\mathrm{XSL(\rho,\gamma,\delta)}=\hat{S}^\dagger(\ln\rho)\mathrm{XSL(1,\gamma,\delta)}\hat{S}(\ln\rho).
\eea
Here, $\hat{S}(z)\equiv e^{\frac{1}{2}(z^*\hat{a}^2-z\hat{a}^{\dagger 2})}$ is the squeezing operator that satisfies $\hat{S}^\dagger(z)\hat{\varphi}\hat{S}(z)=e^{-z}\hat{\varphi}$ for a real parameter $z\in\mathbb{R}$. The squeezing operation is a standard Gaussian operation that has been realized in superconducting circuits with a squeezed drive \cite{puri2017npjq,ma2021sb,eriksson2024nc}.

\begin{figure}[ptb]
\centering
\includegraphics[width=0.9\linewidth]{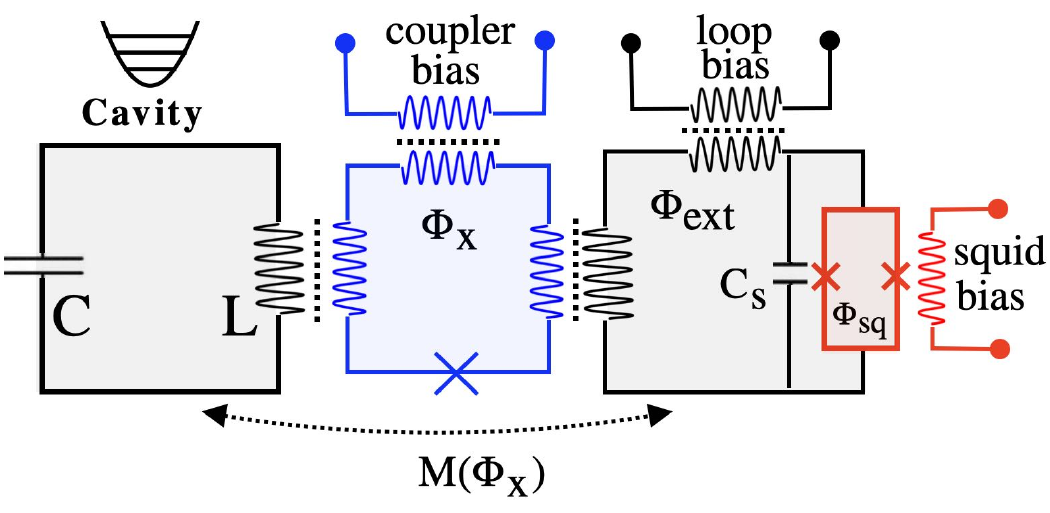}
\caption{ Experimental implementation. A superconducting cavity (LC circuit, left) coupled to a SQUID loop (red circuit, right) shunted by a large capacitance $C_s$ via a coupler (blue circuit, middle). $M(\Phi_x)$ is the effective mutual inductance between the cavity and SQUID circuit induced by the coupler. $\Phi_x$, $\Phi_{ext}$ and $\Phi_{sq}$ are the net fluxes applied to the coupler, SQUID external loop and SQUID itself,  respectively, and can be tuned via the coupler bias, loop bias and squid bias lines. }
\label{Fig-Exps}
\end{figure}

Alternatively, one can implement the quantum lattice gate operations by introducing inductive coupling between the cavity and the JJ-based loops. As illustrated in Fig.~\ref{Fig-Exps},  a superconducting cavity (left) couples to, via a mediated JJ-based circuit (middle) referred to as the ``coupler", the SQUID device (right) shunted by a large capacitance $C_s$ with an external loop. The coupler induces a tunable effective mutual inductance $M(\Phi_x)$ between the cavity and the SQUID circuit \cite{allman2010prl}. The net flux $\Phi_x$, $\Phi_{ext}$ and $\Phi_{sq}$ that applied to the coupler, the SQUID external loop and the SQUID itself can be tuned via the coupler bias line, loop bias line and SQUID bias line respectively.  %
In total, the dynamics of superconducting cavity is  described by the Hamiltonian 
$  \hat{\mathcal{H}}(t)=\hbar\omega_0\hat{a}^\dagger\hat{a}+V_{\mathrm{SQUID}}(\hat{\varphi})$ with $\omega_0=1/\sqrt{LC}$ the bare frequency of cavity. Note that we have assumed that the charging energy of SQUID circuit is suppressed by the large shunted capacitance $C_s$ \cite{koch2007pra}. The inductive energy of SQUID circuit is 
 \bea\label{eq-squid}
V_{\mathrm{SQUID}}(\hat{\varphi})=-E_J\cos(\frac{\Phi_{sq}}{\Phi_0}\pi)\cos\Big[\frac{M(\Phi_x)}{L}\hat{\varphi}+\frac{\Phi_{ext}}{\Phi_0}\Big],\ \ \ \  \ \ 
 \eea
  where $E_J$ is the single JJ energy, $\Phi_0=2\pi\hbar/2e$ is the magnetic flux quantum and $\hat{\varphi}=\sqrt{\frac{\lambda}{2}}(\hat{a}^\dagger+\hat{a})=2\pi\hat{\Phi}/\Phi_0$ with dimensionless Planck constant $\lambda=4e^2\omega_0L/\hbar$ is the cavity flux variable. 
  The effective mutual inductance can be tuned, via the coupler bias, to be zero [$M(\Phi_x)=0$], arbitrarily large [$M(\Phi_x)\rightarrow\infty$], or even negative [$M(\Phi_x)<0]$~\cite{allman2010prl}. By turn on/off rapidly the coupling between the cavity and the SQUID circuit via the mutual inductance or the Josephson energy of SQUID, one can in principle implement the lattice gate
$
\mathrm{XSL(\rho,\gamma,\delta)}= \exp[i\gamma\cos(\rho\hat{\varphi}+\delta)],
$
cf. Eq.~(\ref{eq-XSL}),
where the gate parameters $\gamma$, $\rho$ and $\delta$ are tuned by the biased flux $\Phi_{sq}$, $\Phi_{x}$ and $\Phi_{ext}$ respectively.

\subsection{Summary and Outlook}\label{sec-sum}

In this work, we have introduced a universal quantum lattice gate for bosonic quantum computation, which can be implemented in a JJ-based superconducting circuit architecture. Compared to other popular gates, such as the SNAP gate~\cite{Krastanov2015pra} and the cubic phase gate~\cite{lloyd1999prl,sefi2011prl}, which typically overlook higher-order nonlinearities in the JJ potential, our quantum lattice gates leverage the full nonlinearity of the JJ circuit as a powerful quantum resource. We have developed a comprehensive analytical framework to decompose a given bosonic code engineering process into sequences of quantum lattice gates through Floquet Hamiltonian engineering. Our proposal also offers a direct method for constructing target Hamiltonians aligned precisely with the desired target codes. We have demonstrated the versatility of our method across various applications, including code state preparation, code space embedding, and code space transformation, specifically with cat and binomial code states. We have also studied the automatic quantum error correction of cat code states against photon loss errors with our driving protocol. The results not only advance the toolkit for bosonic quantum computation, but also pave the way for more robust and scalable implementations in superconducting quantum circuits.

In general, our lattice gate decomposition for code state engineering is based on the adiabatic ramp protocol~\cite{xanda2023arxiv}. In the future, we plan to explore methods to accelerate  state engineering by incorporating counter-adiabatic terms, following Berry's solution~\cite{berry2009jpa}. These terms can be engineered through additional driving fields using our technique of arbitrary Floquet Hamiltonian engineering~\cite{guo2024prl}, or by directly optimizing the sequence and parameters of the quantum lattice gates with various numerical methods~\cite{fosel2020arxiv,khaneia2005jmr,fosel2018prx}. In short, it would be promising to enhance the efficiency and speed of our state engineering protocols, pushing the boundaries of what can be achieved in bosonic quantum computation.

\bigskip

\textbf{Acknowledgements}

 
We acknowledge helpful discussions with Vittorio Peano. This work was supported by the National Natural Science Foundation of China (Grant No. 12475025). L.D. acknowledges financial support by the Knut och Alice Wallenberg stiftelse through project grant no. 2022.0090. T.Y.H acknowledges financial support by the Knut and Alice Wallenberg through the Wallenberg Center for Quantum Technology (WACQT).



\onecolumngrid

\appendix

\section{Wigner functions and noncommutative Fourier coefficients}\label{app-Fourier}

In this appendix section, we demonstrate how to derive the Wigner function of a given target state and the noncommutative Fourier coefficients of the corresponding target Hamiltonian.

\subsection{Wigner functions of arbitrary target states}

By expressing a given quantum state $|\psi_{T}\rangle$ as a superposition of Fock states, i.e., $|\psi_{T}\rangle=\sum_{m}c_{m}|m\rangle$, the corresponding density operator can be written as $\hat{\rho}_{T}=\sum_{m,m'}\rho_{m,m'}|m\rangle\langle m'|$ with $\rho_{m,m'}=\langle m|\hat{\rho}|m'\rangle$. The Wigner function of each component $|m\rangle\langle m'|$ can be readily obtained as
\begin{equation}
\begin{split}
W_{mm'}&=\frac{1}{\pi^{2}}\int e^{\xi^{*}\alpha-\xi\alpha^{*}}\text{Tr}\left[|m\rangle\langle m'|e^{\xi \hat{a}^{\dag}-\xi^{*}\hat{a}}\right]d\xi^{2} \\
&=(-1)^{m'}\frac{2}{\pi}\sqrt{\frac{m'!}{m!}}e^{-2|\alpha|^{2}}(4|\alpha|^{2})^{-(m'-m)/2}L_{m'}^{m-m'}(4|\alpha|^{2})e^{i(m'-m)\text{Arg}(\alpha)},
\end{split}
\label{Laguerre}
\end{equation}
where $L_{m'}^{m-m'}(x)=\frac{1}{m'!}e^{x}x^{m'-m}\frac{\partial^{m'}}{\partial x^{m'}}x^{m}e^{-x}$ is the generalized Laguerre polynomial, satisfying the relation $L_{l}^{n}(z)/L_{l+n}^{-n}=(-z)^{-n}(l+n)!/j!$ for $z>0$. The entire Wigner function of the state is given by 
\begin{equation}
W(\alpha)=\sum_{m,m'}\rho_{m,m'}W_{m,m'}.
\label{overallW}
\end{equation}
As an example, the Wigner functions of the binomial code states  $\hat{\rho}_{b,0}=|\bar{0}_{b}\rangle\langle \bar{0}_{b}|$ and $\hat{\rho}_{b,1}=|\bar{1}_{b}\rangle\langle \bar{1}_{b}|$ can be obtained as 
\begin{eqnarray}
W_{b}^{0}(\alpha)&=&\frac{1}{4}\left[W_{00}+3W_{44}+\sqrt{3}(W_{04}+W_{40})\right], \label{wignerb0 }\\
W_{b}^{1}(\alpha)&=&\frac{1}{4}\left[3W_{22}+W_{66}+\sqrt{3}(W_{26}+W_{62})\right]. \label{wignerb1}
\end{eqnarray}
For the four-legged cat code states
$\hat{\rho}_{c,0}=|\bar{0}_{c}\rangle\langle \bar{0}_{c}|$ and $\hat{\rho}_{c,1}=|\bar{1}_{c}\rangle\langle \bar{1}_{c}|$, the Wigner functions are given by
\begin{eqnarray}
W_{c}^{0}(\alpha)&=&\frac{16}{\mathcal{N}_{0}}e^{-\alpha^{2}}\sum_{m,m'}\frac{\alpha^{4(m+m'})}{\sqrt{(4m)!(4m')!}}W_{4m,4m'}, \label{wignerc0 }\\
W_{c}^{1}(\alpha)&=&\frac{16}{\mathcal{N}_{2}}e^{-\alpha^{2}}\sum_{m,m'}\frac{\alpha^{4(m+m'+1)}}{\sqrt{(4m+2)!(4m'+2)!}}W_{4m+2,4m'+2}. \label{wignerc1}
\end{eqnarray}

\subsection{Noncommutative Fourier coefficients of target states and Hamiltonians}

For an arbitrary target Hamiltonian written in the form $\hat{H}_{T}=\sum_{m,m'}c_{m,m'}|m\rangle\langle m'|$, where $c_{m,m'}$ is the superposition coefficient determined by the target logical states, its Husimi Q-function can be readily obtained as
\begin{eqnarray}
H^{Q}_{T}(x,p)&=&\langle\alpha|\hat{H}^{T}|\alpha\rangle \nonumber\\
&=&e^{-|\alpha|^{2}}\sum_{m,m'}\sum_{n,n'}c_{m,m'}\frac{(\alpha^{*})^{n}\alpha^{n'}}{\sqrt{n!n'!}}\langle n|m\rangle\langle m'|n'\rangle \nonumber\\
&=&e^{-|\alpha|^{2}}\sum_{m,m'}c_{m,m'}\frac{(\alpha^{*})^{m}\alpha^{m'}}{\sqrt{m!m'!}} \nonumber\\
&=&e^{-\frac{x^{2}+p^{2}}{2\lambda}}\sum_{m,m'}c_{m,m'}\frac{\left(\frac{x-ip}{\sqrt{2\lambda}}\right)^{m}\left(\frac{x+ip}{\sqrt{2\lambda}}\right)^{m'}}{\sqrt{m!m'!}},
\label{HusimiQ}
\end{eqnarray}
where $|\alpha\rangle$ is the coherent state defined as $\hat{a}|\alpha\rangle=\alpha|\alpha\rangle$, with $\alpha=\left(x+ip\right)/\sqrt{2\lambda}$.
By transforming into polar coordinates with $x=r\cos{\theta}$ and $p=r\sin{\theta}$, and introducing $k_{x}=k\cos{\Omega t}$ and $k_{p}=k\sin{\Omega t}$ as in the main text, the noncommutative Fourier coefficient of the target Hamiltonian, cf. Eq.~(\ref{eq-fFT}) in the main text, can be expressed as
\begin{equation}
f_{T}(k,\tau)=\frac{e^{\frac{\lambda}{4}k^{2}}}{2\pi}\int\int rdrd\theta H^{T}_{Q}(r\cos{\theta},r\sin{\theta})e^{-ikr\cos{(\theta-\tau)}}
\label{fTdef}
\end{equation} 
with $\tau=\Omega t$. In this way, each component of the noncommutative Fourier coefficient can be derived as
\begin{eqnarray}
f_{mm'}(k,\tau)&=&\langle m|f_{T}(k,\tau)|m'\rangle \nonumber\\
&=&\frac{e^{\frac{\lambda}{4}k^{2}}}{2\pi}\int_{0}^{+\infty}rdr\int_{0}^{2\pi}d\theta\langle m|H^{T}_{Q}(r\cos{\theta},r\sin{\theta})|m'\rangle e^{-ikr\cos{(\theta-\tau)}} \nonumber\\
&=&\frac{e^{\frac{\lambda}{4}k^{2}}}{2\pi}\frac{e^{i(m'-m)\tau}}{\sqrt{m!m'!}}\left(\frac{1}{\sqrt{2\lambda}}\right)^{m+m'}\int_{0}^{+\infty}r^{m+m'+1}e^{-\frac{r^{2}}{2\lambda}}dr\int_{0}^{2\pi}e^{-ikr\cos{(\theta-\tau)}}e^{i(m'-m)(\theta-\tau)}d\theta \nonumber\\
&=&e^{\frac{\lambda}{4}k^{2}}\frac{e^{i(m'-m)\tau}}{\sqrt{m!m'!}}i^{m-m'}\left(\frac{1}{\sqrt{2\lambda}}\right)^{m+m'}\int_{0}^{+\infty}r^{m+m'+1}e^{-\frac{r^{2}}{2\lambda}}J_{m-m'}(-kr)dr \nonumber\\
&=&\left \{ \begin{array}{lll}
e^{\frac{\lambda}{4}k^{2}}\sqrt{\frac{m!}{m'!}}\left(ie^{i\tau}\frac{1}{k}\sqrt{\frac{2}{\lambda}}\right)^{m'-m}\frac{\lambda}{\Gamma(1+m-m')}{}_{1}F_{1}(1+m; 1+m-m'; -\frac{\lambda}{2}k^{2}), \quad m-m'>-1, \\
e^{\frac{\lambda}{4}k^{2}}\sqrt{\frac{m'!}{m!}}\left(ie^{-i\tau}\frac{1}{k}\sqrt{\frac{2}{\lambda}}\right)^{m-m'}\frac{\lambda}{\Gamma(1+m'-m)}{}_{1}F_{1}(1+m'; 1+m'-m; -\frac{\lambda}{2}k^{2}), \quad m-m'\leq -1.
\end{array} \right.
\label{fTcomponent}
\end{eqnarray} 
Here, $J_{m-m'}(z)$ is the Bessel function of the first kind (of order $m-m'$) satisfying the identity
\begin{equation}
J_{z}(x)=\frac{i^{-z}}{2\pi}\int_{0}^{2\pi}e^{i(zy+x\cos{y})}dy,
\end{equation}
$\Gamma(n)=(n-1)!$ is the Gamma function, and ${}_{1}F_{1}(a;b;z)$ is the Kummer confluent hypergeometric function. Note that, in the final step of Eq.~(\ref{fTcomponent}), we have classified the identity into two different situations for the convenience of numerical simulations. The entire noncommutative Fourier coefficient can be immediately obtained as
\begin{equation}
f_{T}(k,\tau)=\sum_{m,m'}c_{m,m'}f_{m,m'}(k,\tau).
\label{fTcomplete}
\end{equation}

Following the above procedure, the noncommutative Fourier coefficient of the target Hamiltonian $\hat{H}_T=-\Delta|\psi_{T}\rangle\langle \psi_{T} |$, with the target state $|\psi_{T}\rangle$ given by Eq.~(\ref{superposition}) in the main text, is obtained as 
\begin{eqnarray}
f_{T}(k,\tau)&=&-\frac{\Delta}{4}\Big[|c_{0}|^{2}f_{00}+3|c_{1}|^{2}f_{22}+3|c_{0}|^{2}f_{44}+|c_{1}|^{2}f_{66}\nonumber\\
&&+\sqrt{3}\left(|c_{0}|^{2}f_{04}+|c_{1}|^{2}f_{26}+c_{0}c_{1}^{*}f_{02}+c_{0}c_{1}^{*}f_{46}+\text{c.c.}\right)+\left(3c_{1}c_{0}^{*}f_{24}+c_{0}c_{1}^{*}f_{06}+\text{c.c.}\right)\Big],
\label{appfT1}
\end{eqnarray}
where we have used the relation $f_{mm'}=(f_{m'm})^{*}$ that is valid if $|m-m'|$ is even. Similarly, the noncommutative Fourier coefficient of the target Hamiltonian given by Eq.~(\ref{eq-HTembedbin}) is 
\begin{equation}
f_{T}(k,\tau)=-\frac{\Delta}{4}\left[f_{00}+3f_{22}+3f_{44}+f_{66}+\sqrt{3}\left(f_{04}+f_{26}+\text{c.c.}\right)\right].
\label{appfT2}
\end{equation}
Once again, one has $f_{mm'}=(f_{m'm})^{*}$ since $m$ and $m'$ are always even integers.

\section{Noncommutative Fourier coefficient for the code states transformation}\label{app-cst}

In this appendix, we derive the analytical expression of the noncommutative Fourier coefficient for the code-space transformation process discussed in Sec.~\ref{sec-cst}. 
According to Eq.~(\ref{eq-transt}), the target Hamiltonian given by Eq.~(\ref{eq-Htran}) can be expressed as a sum of three contributions, i.e., $\hat{H}_{T}(t)=\hat{H}_{b,t}+\hat{H}_{c,t}+H_{\text{cross},t}$, with
\begin{eqnarray}
\hat{H}_{b,t}&=&-\Delta[1-h(t)]\left(|\bar{0}_{b}\rangle\langle\bar{0}_{b}|+|\bar{1}_{b}\rangle\langle\bar{1}_{b}|\right) \label{appHbt}\\
\hat{H}_{c,t}&=&-\Delta h(t)\left(|\bar{0}_{c}\rangle\langle\bar{0}_{c}|+|\bar{1}_{c}\rangle\langle\bar{1}_{c}|\right) \label{appHct}\\
\hat{H}_{\text{cross},t}&=&-\Delta\sqrt{h(t)[1-h(t)]}\left(|\bar{0}_{b}\rangle\langle\bar{0}_{c}|+|\bar{1}_{b}\rangle\langle\bar{1}_{c}|+\text{H.c.}\right). \label{appHcrosst}
\end{eqnarray}
The first two terms involve only the binomial and cat code states, respectively, while the last term contains the cross terms between them.

The noncommutative Fourier components of $\hat{H}_{b,t}$ can be straightforwardly obtained from the results in Eq.~(\ref{appfT2}), with consideration of the time-dependent coefficient $1-h(t)$, i.e.,
\begin{equation}
f_{b,T}(k,\tau)=-\frac{\Delta[1-h(t)]}{4}\left[f_{00}+3f_{22}+3f_{44}+f_{66}+\sqrt{3}\left(f_{04}+f_{26}+\text{c.c.}\right)\right].
\label{fbt}
\end{equation}

To calculate the noncommutative Fourier coefficient of $\hat{H}_{c,t}$, which comprises only the four-legged cat states given in Eq.~(\ref{eq-sm-4cat2}), we first calculate its Husimi-Q function as follows:
\begin{eqnarray}
H^{Q}_{c}(x,p)&=&\langle\alpha|\hat{H}_{c,t}|\alpha\rangle \nonumber\\
&=&-\Delta h(t)\left(\big|\langle\alpha|\bar{0}_{c}\rangle\big|^{2}+\big|\langle\alpha|\bar{1}_{c}\rangle\big|^{2}\right) \nonumber\\
&=&-16\Delta h(t)e^{-|\alpha|^{2}}e^{-\tilde{\alpha}^{2}}\left(\frac{1}{\mathcal{N}_{0}}\sum_{n,m=0}^{\infty}\frac{(\alpha^{*})^{4n}\alpha^{4m}\tilde{\alpha}^{4(n+m)}}{(4n)!(4m)!}+\frac{1}{\mathcal{N}_{2}}\sum_{n,m=0}^{\infty}\frac{(\alpha^{*})^{4n+2}\alpha^{4m+2}\tilde{\alpha}^{4(n+m)+4}}{(4n+2)!(4m+2)!}\right) \nonumber\\
&=&-16\Delta h(t)e^{-\frac{x^{2}+p^{2}}{2\lambda}}e^{-\tilde{\alpha}^{2}}\left[\frac{1}{\mathcal{N}_{0}}\sum_{n,m}^{\infty}\frac{\tilde{\alpha}^{4(n+m)}}{(4n)!(4m)!}\left(\frac{x-ip}{\sqrt{2\lambda}}\right)^{4n}\left(\frac{x+ip}{\sqrt{2\lambda}}\right)^{4m}\right.\nonumber\\
&&\left.+\frac{1}{\mathcal{N}_{2}}\sum_{n,m}^{\infty}\frac{\tilde{\alpha}^{4(n+m)+4}}{(4n+2)!(4m+2)!}\left(\frac{x-ip}{\sqrt{2\lambda}}\right)^{4n+2}\left(\frac{x+ip}{\sqrt{2\lambda}}\right)^{4m+2}\right]
\label{Qcat}
\end{eqnarray} 
with $\tilde{\alpha}$ the amplitude of each component of the cat states. Again, using the method in Appendix~\ref{app-Fourier}, the noncommutative Fourier coefficient of $H_{c}^{Q}$ can be straightforwardly obtained as 
\begin{eqnarray}
f_{c,T}(k,\tau)&=&\frac{e^{\frac{\lambda k^{2}}{4}}}{2\pi}\int\int rdrd\theta H^{Q}_{c}(r\cos{\theta},r\sin{\theta})e^{-ikr\cos{(\theta-\tau)}} \nonumber\\
&=&\frac{e^{\frac{\lambda k^{2}}{4}}}{2\pi}\left[\mathcal{M}_{0}(t)\sum_{n,m=0}^{\infty}\frac{e^{4i(m-n)\tau}\tilde{\alpha}^{4(n+m)}}{(4n)!(4m)!}\left(\frac{1}{2\lambda}\right)^{2(n+m)}\int_{0}^{\infty}r^{4(n+m)+1}e^{-\frac{r^{2}}{2\lambda}}dr\int_{0}^{2\pi} e^{-ikr\cos{(\theta-\tau)}}e^{4i(m-n)(\theta-\tau)}d\theta \right.\nonumber\\
&&\left.+\mathcal{M}_{2}(t)\sum_{n,m=0}^{\infty}\frac{e^{4i(m-n)\tau}\tilde{\alpha}^{4(n+m)+4}}{(4n+2)!(4m+2)!}\left(\frac{1}{2\lambda}\right)^{2(n+m)+2}\int_{0}^{\infty}r^{4(n+m)+5}e^{-\frac{r^{2}}{2\lambda}}dr\int_{0}^{2\pi} e^{-ikr\cos{(\theta-\tau)}}e^{4i(m-n)(\theta-\tau)}d\theta\right] \nonumber\\
&=&e^{\frac{\lambda k^{2}}{4}}\left[\mathcal{M}_{0}(t)\sum_{n,m=0}^{\infty}\frac{e^{4i(m-n)\tau}\tilde{\alpha}^{4(n+m)}}{(4n)!(4m)!}i^{4(n-m)}\left(\frac{1}{2\lambda}\right)^{2(n+m)}\int_{0}^{\infty}r^{4(n+m)+1}e^{-\frac{r^{2}}{2\lambda}}J_{4(n-m)}(-kr)dr\right. \nonumber\\
&&\left.+\mathcal{M}_{2}(t)\sum_{n,m=0}^{\infty}\frac{e^{4i(m-n)\tau}\tilde{\alpha}^{4(n+m)+4}}{(4n+2)!(4m+2)!}i^{4(n-m)}\left(\frac{1}{2\lambda}\right)^{2(n+m)+2}\int_{0}^{\infty}r^{4(n+m)+5}e^{-\frac{r^{2}}{2\lambda}}J_{4(n-m)}(-kr)dr\right] \nonumber\\
&=&\sum_{n,m=0}^{\infty}\left[\mathcal{M}_{0}(t)\frac{\tilde{\alpha}^{4(n+m)}}{\sqrt{(4n)!(4m)!}}f_{4n,4m}+\mathcal{M}_{2}(t)\frac{\tilde{\alpha}^{4(n+m)+4}}{\sqrt{(4n+2)!(4m+2)!}}f_{4n+2,4m+2}\right].
\label{fcc}
\end{eqnarray}
with $\mathcal{M}_{0(2)}(t)=-16\Delta h(t)\text{exp}(-\tilde{\alpha}^{2})/\mathcal{N}_{0(2)}$ and $f_{mm'}$ given in Eq.~(\ref{fTcomponent}).

For the cross-part Hamiltonian $\hat{H}_{\text{cross},t}$, we first expand it as follows:
\begin{eqnarray}
\hat{H}_{\text{cross},t}
&=&-\Delta \sqrt{[1-h(t)]h(t)}(|\bar{0}_{b}\rangle\langle \bar{0}_{c}|+|\bar{1}_{b}\rangle\langle \bar{1}_{c}|)+\textrm{H.c.} \nonumber\\
&=&-2\Delta e^{-\frac{\tilde{\alpha}^{2}}{2}}\sqrt{[1-h(t)]h(t)}\left[\frac{1}{\sqrt{\mathcal{N}_{0}}}\sum_{n=0}^{\infty} \frac{\tilde{\alpha}^{4n}}{\sqrt{(4n)!}}\left(|0\rangle+\sqrt{3}|4\rangle\right)\langle 4n| \right.\nonumber\\
&&\left.+\frac{1}{\sqrt{\mathcal{N}_{2}}}\sum_{n=0}^{\infty} \frac{\tilde{\alpha}^{4n+2}}{\sqrt{(4n+2)!}}\left(\sqrt{3}|2\rangle+|6\rangle\right)\langle 4n+2|+\textrm{H.c.}\right],
\label{bcHT}
\end{eqnarray}
and then calculate its Husimi-Q function $$H_{\text{cross}}^{Q}=\langle\alpha|\hat{H}_{\text{cross},t}|\alpha\rangle=-2\Delta e^{-\frac{\tilde{\alpha}^{2}}{2}}\sqrt{[1-h(t)]h(t)}\left(H^{Q}_{0}+H^{Q}_{2}+H^{Q}_{4}+H^{Q}_{6}+\text{c.c.}\right),$$ 
where
\bea
H^{Q}_{0}(x,p)&=&\frac{1}{\sqrt{\mathcal{N}_{0}}}e^{-|\alpha|^{2}}\sum_{n=0}^{\infty} \frac{\left(\alpha\tilde{\alpha}\right)^{4n}}{(4n)!} 
=\frac{1}{\sqrt{\mathcal{N}_{0}}}e^{-\frac{x^{2}+p^{2}}{2\lambda}}\sum_{n=0}^{\infty}\frac{\tilde{\alpha}^{4n}}{(4n)!}\left(\frac{x+ip}{\sqrt{2\lambda}}\right)^{4n}, \\
H^{Q}_{2}(x,p)&=&\frac{\sqrt{3}}{\sqrt{\mathcal{N}_{2}}}e^{-|\alpha|^{2}}\sum_{n=0}^{\infty} \frac{\left(\alpha^{*}\right)^{2}\left(\alpha\tilde{\alpha}\right)^{4n+2}}{\sqrt{2!}(4n+2)!} =\frac{\sqrt{3}}{\sqrt{\mathcal{N}_{1}}}e^{-\frac{x^{2}+p^{2}}{2\lambda}}\sum_{n=0}^{\infty}\frac{\tilde{\alpha}^{4n+2}}{\sqrt{2!}(4n+2)!}\left(\frac{x^{2}+p^{2}}{2\lambda}\right)^{2}\left(\frac{x+ip}{\sqrt{2\lambda}}\right)^{4n}, \ \ \ \ \ \ \\
H^{Q}_{4}(x,p)&=&\frac{\sqrt{3}}{\sqrt{\mathcal{N}_{0}}}e^{-|\alpha|^{2}}\sum_{n=0}^{\infty} \frac{\left(\alpha^{*}\right)^{4}\left(\alpha\tilde{\alpha}\right)^{4n}}{\sqrt{4!}(4n)!}=\frac{\sqrt{3}}{\sqrt{\mathcal{N}_{0}}}e^{-\frac{x^{2}+p^{2}}{2\lambda}}\sum_{n=0}^{\infty}\frac{\tilde{\alpha}^{4n}}{\sqrt{4!}(4n)!}\left(\frac{x^{2}+p^{2}}{2\lambda}\right)^{4}\left(\frac{x+ip}{\sqrt{2\lambda}}\right)^{4n-4}, \\
H^{Q}_{6}(x,p)&=&\frac{1}{\sqrt{\mathcal{N}_{2}}}e^{-|\alpha|^{2}}\sum_{n=0}^{\infty} \frac{\left(\alpha^{*}\right)^{6}\left(\alpha\tilde{\alpha}\right)^{4n+2}}{\sqrt{6!}(4n+2)!} =\frac{1}{\sqrt{\mathcal{N}_{2}}}e^{-\frac{x^{2}+p^{2}}{2\lambda}}\sum_{n=0}^{\infty}\frac{\tilde{\alpha}^{4n+2}}{\sqrt{6!}(4n+2)!}\left(\frac{x^{2}+p^{2}}{2\lambda}\right)^{6}\left(\frac{x+ip}{\sqrt{2\lambda}}\right)^{4n-4},
\eea
with $\alpha=(x+ip)/\sqrt{2\lambda}$. The noncommutative Fourier components of the above four parts can be obtained as follows:
\bea
f_{\text{cross},0}(k,\tau)&=&\frac{e^{\frac{\lambda k^{2}}{4}}}{2\pi}\int\int rdrd\theta H^{Q}_{0}(r\cos{\theta},r\sin{\theta})e^{-ikr\cos{(\theta-\tau)}} \nonumber\\
&=&\frac{e^{\frac{\lambda k^{2}}{4}}}{2\pi\sqrt{\mathcal{N}_{0}}}\sum_{n=0}^{\infty}\frac{e^{4in\tau}\tilde{\alpha}^{4n}}{(4n)!}\left(\frac{1}{2\lambda}\right)^{2n}\int_{0}^{\infty}r^{4n+1}e^{-\frac{r^{2}}{2\lambda}}dr\int_{0}^{2\pi} e^{-ikr\cos{(\theta-\tau)}}e^{4in(\theta-\tau)}d\theta \nonumber\\
&=&\frac{e^{\frac{\lambda k^{2}}{4}}}{\sqrt{\mathcal{N}_{0}}}\sum_{n=0}^{\infty}\frac{e^{4in\tau}\tilde{\alpha}^{4n}}{(4n)!}i^{-4n}\left(\frac{1}{2\lambda}\right)^{2n}\int_{0}^{\infty}r^{4n+1}e^{-\frac{r^{2}}{2\lambda}}J_{-4n}(-kr)dr \nonumber\\
&=&\frac{1}{\sqrt{\mathcal{N}_{0}}}\sum_{n=0}^{\infty}\frac{\tilde{\alpha}^{4n}}{\sqrt{(4n)!}}f_{0,4n}, \label{fc0} 
\eea
\bea
f_{\text{cross},2}(k,\tau)&=&\frac{e^{\frac{\lambda k^{2}}{4}}}{2\pi}\int\int rdrd\theta H^{Q}_{2}(r\cos{\theta},r\sin{\theta})e^{-ikr\cos{(\theta-\tau)}} \nonumber\\
&=&\frac{e^{\frac{\lambda k^{2}}{4}}\sqrt{3}}{2\pi\sqrt{\mathcal{N}_{2}}}\sum_{n=0}^{\infty}\frac{e^{4in\tau}\tilde{\alpha}^{4n+2}}{\sqrt{2!}(4n+2)!}\left(\frac{1}{2\lambda}\right)^{2(n+1)}\int_{0}^{\infty}r^{4(n+1)+1}e^{-\frac{r^{2}}{2\lambda}}dr\int_{0}^{2\pi} e^{-ikr\cos{(\theta-\tau)}}e^{4in(\theta-\tau)}d\theta \nonumber\\
&=&\frac{e^{\frac{\lambda k^{2}}{4}}\sqrt{3}}{\sqrt{\mathcal{N}_{2}}}\sum_{n=0}^{\infty}\frac{e^{4in\tau}\tilde{\alpha}^{4n+2}}{\sqrt{2!}(4n+2)!}i^{-4n}\left(\frac{1}{2\lambda}\right)^{2(n+1)}\int_{0}^{\infty}r^{4(n+1)+1}e^{-\frac{r^{2}}{2\lambda}}J_{-4n}(-kr)dr \nonumber\\
&=&\frac{\sqrt{3}}{\sqrt{\mathcal{N}_{2}}}\sum_{n=0}^{\infty}\frac{\sqrt{2!}\tilde{\alpha}^{4n+2}}{\sqrt{(4n+2)!}}f_{2,4n+2}, \label{fc2} 
\eea
\bea
f_{\text{cross},4}(k,\tau)&=&\frac{e^{\frac{\lambda k^{2}}{4}}}{2\pi}\int\int rdrd\theta H^{Q}_{4}(r\cos{\theta},r\sin{\theta})e^{-ikr\cos{(\theta-\tau)}} \nonumber\\
&=&\frac{e^{\frac{\lambda k^{2}}{4}}\sqrt{3}}{2\pi\sqrt{\mathcal{N}_{0}}}\sum_{n=0}^{\infty}\frac{e^{4i(n-1)\tau}\tilde{\alpha}^{4n}}{\sqrt{4!}(4n)!}\left(\frac{1}{2\lambda}\right)^{2(n+1)}\int_{0}^{\infty}r^{4(n+1)+1}e^{-\frac{r^{2}}{2\lambda}}dr\int_{0}^{2\pi} e^{-ikr\cos{(\theta-\tau)}}e^{4i(n-1)(\theta-\tau)}d\theta \nonumber\\
&=&\frac{e^{\frac{\lambda k^{2}}{4}}\sqrt{3}}{\sqrt{\mathcal{N}_{0}}}\sum_{n=0}^{\infty}\frac{e^{4i(n-1)\tau}\tilde{\alpha}^{4n}}{\sqrt{4!}(4n)!}i^{-4(n-1)}\left(\frac{1}{2\lambda}\right)^{2(n+1)}\int_{0}^{\infty}r^{4(n+1)+1}e^{-\frac{r^{2}}{2\lambda}}J_{-4(n-1)}(-kr)dr \nonumber\\
&=&\frac{\sqrt{3}}{\sqrt{\mathcal{N}_{0}}}\sum_{n=0}^{\infty}\frac{\sqrt{4!}\tilde{\alpha}^{4n}}{\sqrt{(4n)!}}f_{4,4n}, \label{fc4} 
\eea
\bea
f_{\text{cross},6}(k,\tau)&=&\frac{e^{\frac{\lambda k^{2}}{4}}}{2\pi}\int\int rdrd\theta H^{Q}_{6}(r\cos{\theta},r\sin{\theta})e^{-ikr\cos{(\theta-\tau)}} \nonumber\\
&=&\frac{e^{\frac{\lambda k^{2}}{4}}}{2\pi\sqrt{\mathcal{N}_{2}}}\sum_{n=0}^{\infty}\frac{e^{4i(n-1)\tau}\tilde{\alpha}^{4n+2}}{\sqrt{6!}(4n+2)!}\left(\frac{1}{2\lambda}\right)^{2(n+2)}\int_{0}^{\infty}r^{4(n+2)+1}e^{-\frac{r^{2}}{2\lambda}}dr\int_{0}^{2\pi} e^{-ikr\cos{(\theta-\tau)}}e^{4i(n-1)(\theta-\tau)}d\theta \nonumber\\
&=&\frac{e^{\frac{\lambda k^{2}}{4}}}{\sqrt{\mathcal{N}_{2}}}\sum_{n=0}^{\infty}\frac{e^{4i(n-1)\tau}\tilde{\alpha}^{4n+2}}{\sqrt{6!}(4n+2)!}i^{-4(n-1)}\left(\frac{1}{2\lambda}\right)^{2(n+2)}\int_{0}^{\infty}r^{4(n+2)+1}e^{-\frac{r^{2}}{2\lambda}}J_{-4(n-1)}(-kr)dr \nonumber\\
&=&\frac{1}{\sqrt{\mathcal{N}_{2}}}\sum_{n=0}^{\infty}\frac{\sqrt{6!}\tilde{\alpha}^{4n+2}}{\sqrt{(4n+2)!}}f_{6,4n+2}. \label{fc6}
\eea
As a result, the noncommutative Fourier coefficient for the cross terms is given by
\begin{equation}
f_{\text{cross},T}(k,\tau)=-2\Delta e^{-\frac{\tilde{\alpha}^{2}}{2}}\sqrt{[1-h(t)]h(t)}\left(f_{\text{cross},0}+f_{\text{cross},2}+f_{\text{cross},4}+f_{\text{cross},6}+\text{c.c.}\right).
\label{fcrossT}
\end{equation}

Finally, the entire noncommutative Fourier coefficient $f_{\text{CSpT}}(k,\tau)$ of the total target Hamiltonian involved in the code-space transformation process (CSpT), i.e., $\hat{H}_{T}(t)=\hat{H}_{b,t}+\hat{H}_{c,t}+H_{\text{cross},t}$, is given by the sum of all the previously calculated components, i.e.,
\begin{equation}
f_{\text{CSpT}}(k,\tau)= f_{b,T}(k,\tau)+f_{c,T}(k,\tau)+f_{\text{cross},T}(k,\tau).
\end{equation}

\section{Noncommutative Fourier coefficient for the cat error states}\label{app-error}

In Sec.~\ref{sec-aqec}, we have introduced two cat error states $|\bar{0}_{e}\rangle$ and $|\bar{1}_{e}\rangle$, which are involved in the last two terms of the target Hamiltonian~(\ref{eq-Htran4}).
Following a similar procedure as introduced by Eqs.~(\ref{Qcat}) and (\ref{fcc}), the Husimi-Q function of these two error terms can be calculated as
\begin{eqnarray}
H^{Q}_{e}(x,p)&=&-\Delta h(t)\langle\alpha|\left(|\bar{0}_{e}\rangle\langle\bar{0}_{e}|+|\bar{1}_{e}\rangle\langle\bar{1}_{e}|\right)|\alpha\rangle \nonumber\\
&=&-\Delta h(t)\left(\big|\langle\alpha|\bar{0}_{e}\rangle\big|^{2}+\big|\langle\alpha|\bar{1}_{e}\rangle\big|^{2}\right) \nonumber\\
&=&-16\Delta h(t)e^{-|\alpha|^{2}}e^{-\tilde{\alpha}^{2}}\left(\frac{1}{\mathcal{N}_{1}}\sum_{n,m=0}^{\infty}\frac{(\alpha^{*})^{4n+1}\alpha^{4m+1}\tilde{\alpha}^{4(n+m)+2}}{(4n+1)!(4m+1)!}+\frac{1}{\mathcal{N}_{3}}\sum_{n,m=0}^{\infty}\frac{(\alpha^{*})^{4n+3}\alpha^{4m+3}\tilde{\alpha}^{4(n+m)+6}}{(4n+3)!(4m+3)!}\right) \nonumber\\
&=&-16\Delta h(t)e^{-\frac{x^{2}+p^{2}}{2\lambda}}e^{-\tilde{\alpha}^{2}}\left[\frac{1}{\mathcal{N}_{1}}\sum_{n,m}^{\infty}\frac{\tilde{\alpha}^{4(n+m)+2}}{(4n+1)!(4m+1)!}\left(\frac{x-ip}{\sqrt{2\lambda}}\right)^{4n+1}\left(\frac{x+ip}{\sqrt{2\lambda}}\right)^{4m+1}\right.\nonumber\\
&&\left.+\frac{1}{\mathcal{N}_{3}}\sum_{n,m}^{\infty}\frac{\tilde{\alpha}^{4(n+m)+6}}{(4n+3)!(4m+3)!}\left(\frac{x-ip}{\sqrt{2\lambda}}\right)^{4n+3}\left(\frac{x+ip}{\sqrt{2\lambda}}\right)^{4m+3}\right].
\label{Qerror}
\end{eqnarray} 
The corresponding noncommutative Fourier components are then readily calculated as
\begin{eqnarray}
f_{e,T}(k,\tau)&=&\frac{e^{\frac{\lambda k^{2}}{4}}}{2\pi}\int\int rdrd\theta H^{Q}_{e}(r\cos{\theta},r\sin{\theta})e^{-ikr\cos{(\theta-\tau)}} \nonumber\\
&=&\frac{e^{\frac{\lambda k^{2}}{4}}}{2\pi}\left[\mathcal{M}_{1}(t)\sum_{n,m=0}^{\infty}\frac{e^{4i(m-n)\tau}\tilde{\alpha}^{4(n+m)+2}}{(4n+1)!(4m+1)!}\left(\frac{1}{2\lambda}\right)^{2(n+m)+1}\right.\nonumber\\
&&\left.\times\int_{0}^{\infty}r^{4(n+m)+3}e^{-\frac{r^{2}}{2\lambda}}dr\int_{0}^{2\pi} e^{-ikr\cos{(\theta-\tau)}}e^{4i(m-n)(\theta-\tau)}d\theta\right. \nonumber\\
&&\left.+\mathcal{M}_{3}(t)\sum_{n,m=0}^{\infty}\frac{e^{4i(m-n)\tau}\tilde{\alpha}^{4(n+m)+6}}{(4n+3)!(4m+3)!}\left(\frac{1}{2\lambda}\right)^{2(n+m)+3}\right. \nonumber\\
&&\left.\times\int_{0}^{\infty}r^{4(n+m)+7}e^{-\frac{r^{2}}{2\lambda}}dr\int_{0}^{2\pi} e^{-ikr\cos{(\theta-\tau)}}e^{4i(m-n)(\theta-\tau)}d\theta\right] \nonumber\\
&=&e^{\frac{\lambda k^{2}}{4}}\left[\mathcal{M}_{1}(t)\sum_{n,m=0}^{\infty}\frac{e^{4i(m-n)\tau}\tilde{\alpha}^{4(n+m)+2}}{(4n+1)!(4m+1)!}i^{4(n-m)}\left(\frac{1}{2\lambda}\right)^{2(n+m)+1}\int_{0}^{\infty}r^{4(n+m)+3}e^{-\frac{r^{2}}{2\lambda}}J_{4(n-m)}(-kr)dr\right. \nonumber\\
&&\left.+\mathcal{M}_{3}(t)\sum_{n,m=0}^{\infty}\frac{e^{4i(m-n)\tau}\tilde{\alpha}^{4(n+m)+6}}{(4n+3)!(4m+3)!}i^{4(n-m)}\left(\frac{1}{2\lambda}\right)^{2(n+m)+3}\int_{0}^{\infty}r^{4(n+m)+7}e^{-\frac{r^{2}}{2\lambda}}J_{4(n-m)}(-kr)dr\right] \nonumber\\
&=&\sum_{n,m=0}^{\infty}\left[\mathcal{M}_{1}(t)\frac{\tilde{\alpha}^{4(n+m)+2}}{\sqrt{(4n+1)!(4m+1)!}}f_{4n+1,4m+1}+\mathcal{M}_{3}(t)\frac{\tilde{\alpha}^{4(n+m)+6}}{\sqrt{(4n+3)!(4m+3)!}}f_{4n+3,4m+3}\right],
\label{fcc}
\end{eqnarray}
where $\mathcal{M}_{1(3)}(t)=-16\Delta h(t)\text{exp}(-\tilde{\alpha}^{2})/\mathcal{N}_{1(3)}$ and $f_{mm'}$ is given in Eq.~(\ref{fTcomponent}).

\section{Autonomous quantum error correction via a measurement sequence}\label{app-AQEC}

In this section, we discuss the ability of autonomous quantum error correction (AQEC) against single-photon loss errors using the cat code states with our method.
In fact, the four-legged cat code states $|\bar{0}_{c}\rangle$ and $|\bar{1}_{c}\rangle$ can return to themselves (with only a trivial change in their overall amplitudes) after undergoing \emph{four} single-photon losses:\\
 
 (i) After the first single-photon loss, the code states become 
 \begin{equation}
 \begin{split}
\hat{a}|\hat{0}_{c}\rangle&=\frac{\tilde{\alpha}}{\sqrt{\mathcal{N}_{0}}}\left(|\tilde{\alpha}\rangle-|-\tilde{\alpha}\rangle+i|i\tilde{\alpha}\rangle-i|-i\tilde{\alpha}\rangle\right) =\frac{4\tilde{\alpha}}{\sqrt{\mathcal{N}_{0}}}e^{-\tilde{\alpha}^2/2}\sum_{n}\frac{\tilde{\alpha}^{4n+3}}{\sqrt{(4n+3)!}}|4n+3\rangle \propto |\bar{1}_{e}\rangle.\\
\hat{a}|\hat{1}_{c}\rangle&=\frac{\tilde{\alpha}}{\sqrt{\mathcal{N}_{2}}}\left(|\tilde{\alpha}\rangle-|-\tilde{\alpha}\rangle-i|i\tilde{\alpha}\rangle+i|-i\tilde{\alpha}\rangle\right) =\frac{4\tilde{\alpha}}{\sqrt{\mathcal{N}_{2}}}e^{-\tilde{\alpha}^2/2}\sum_{n}\frac{\tilde{\alpha}^{4n+1}}{\sqrt{(4n+1)!}}|4n+1\rangle \propto |\bar{0}_{e}\rangle.
 \end{split}
 \end{equation}

(ii) After the second single-photon loss, the code states become 
\begin{equation}
 \begin{split}
 \hat{a}^{2}|\bar{0}_{c}\rangle&=\frac{\tilde{\alpha}^{2}}{\sqrt{\mathcal{N}_{0}}}\left(|\tilde{\alpha}\rangle+|-\tilde{\alpha}\rangle-|i\tilde{\alpha}\rangle-|-i\tilde{\alpha}\rangle\right) =\frac{4\tilde{\alpha}^{2}}{\sqrt{\mathcal{N}_{0}}}e^{-\tilde{\alpha}^2/2}\sum_{n}\frac{\tilde{\alpha}^{4n+2}}{\sqrt{(4n+2)!}}|4n+2\rangle \propto |\bar{1}_{c}\rangle.\\
\hat{a}^2|\hat{1}_{c}\rangle&=\frac{\tilde{\alpha}^{2}}{\sqrt{\mathcal{N}_{2}}}\left(|\tilde{\alpha}\rangle+|-\tilde{\alpha}\rangle+|i\tilde{\alpha}\rangle+|-i\tilde{\alpha}\rangle\right) =\frac{4\tilde{\alpha}^{2}}{\sqrt{\mathcal{N}_{2}}}e^{-\tilde{\alpha}^2/2}\sum_{n}\frac{\tilde{\alpha}^{4n}}{\sqrt{(4n)!}}|4n\rangle \propto |\bar{0}_{c}\rangle.
 \end{split}
 \end{equation}

(iii) After the third single-photon loss, the code state become
\begin{equation}
 \begin{split}
 \hat{a}^{3}|\bar{0}_{c}\rangle&=\frac{\tilde{\alpha}^{3}}{\sqrt{\mathcal{N}_{0}}}\left(|\tilde{\alpha}\rangle-|-\tilde{\alpha}\rangle-i|i\tilde{\alpha}\rangle+i|-i\tilde{\alpha}\rangle\right) =\frac{4\tilde{\alpha}^{3}}{\sqrt{\mathcal{N}_{0}}}e^{-\tilde{\alpha}^2/2}\sum_{n}\frac{\tilde{\alpha}^{4n+1}}{\sqrt{(4n+1)!}}|4n+1\rangle \propto |\bar{0}_{e}\rangle.\\
\hat{a}^3|\hat{1}_{c}\rangle&=\frac{\tilde{\alpha}^{3}}{\sqrt{\mathcal{N}_{2}}}\left(|\tilde{\alpha}\rangle-|-\tilde{\alpha}\rangle+i|i\tilde{\alpha}\rangle-i|-i\tilde{\alpha}\rangle\right) =\frac{4\tilde{\alpha}^{3}}{\sqrt{\mathcal{N}_{2}}}e^{-\tilde{\alpha}^2/2}\sum_{n}\frac{\tilde{\alpha}^{4n+3}}{\sqrt{(4n+3)!}}|4n+3\rangle \propto |\bar{1}_{e}\rangle.
 \end{split}
 \end{equation}

(iv) After the fourth single-photon loss, the code states finally become 
\begin{equation}
 \begin{split}
\hat{a}^{4}|\bar{0}_{c}\rangle&=\frac{\tilde{\alpha}^{4}}{\sqrt{\mathcal{N}_{0}}}\left(|\tilde{\alpha}\rangle+|-\tilde{\alpha}\rangle+|i\tilde{\alpha}\rangle+|-i\tilde{\alpha}\rangle\right) =\frac{4\tilde{\alpha}^{4}}{\sqrt{\mathcal{N}_{0}}}e^{-\tilde{\alpha}^2/2}\sum_{n}\frac{\tilde{\alpha}^{4n}}{\sqrt{(4n)!}}|4n\rangle \propto |\bar{0}_{c}\rangle.\\
\hat{a}^4|\hat{1}_{c}\rangle&=\frac{\tilde{\alpha}^{4}}{\sqrt{\mathcal{N}_{2}}}\left(|\tilde{\alpha}\rangle+|-\tilde{\alpha}\rangle-|i\tilde{\alpha}\rangle-|-i\tilde{\alpha}\rangle\right) =\frac{4\tilde{\alpha}^{4}}{\sqrt{\mathcal{N}_{2}}}e^{-\tilde{\alpha}^2/2}\sum_{n}\frac{\tilde{\alpha}^{4n-2}}{\sqrt{(4n-2)!}}|4n-2\rangle \propto |\bar{1}_{c}\rangle.
 \end{split}
 \end{equation}
In short, the four-legged cat code states undergo circulations $|\bar{0}_{c}\rangle\rightarrow|\bar{1}_{e}\rangle\rightarrow|\bar{1}_{c}\rangle\rightarrow|\bar{0}_{e}\rangle\rightarrow|\bar{0}_{c}\rangle$ and $|\bar{1}_{c}\rangle\rightarrow|\bar{0}_{e}\rangle\rightarrow|\bar{0}_{c}\rangle\rightarrow|\bar{1}_{e}\rangle\rightarrow|\bar{1}_{c}\rangle$ under the influence of single-photon loss events.

As discussed in Sec.~\ref{sec-aqec}, to implement the AQEC, we regularly perform the parity measurements defined by
$
\hat{\Pi}_{m}=\sum_{n\in\mathbb{N}^0}|2n+m\rangle\langle 2n+m|
$ with $m=0,1$,
cf. Eq.~(\ref{eq-PIm}) in the main text.
Starting from the initial code state $\hat{\rho}_T(0)=|\psi_c(0)\rangle\langle|\psi_c(0)|$ with
$
|\psi_c(0)\rangle=c_1|\bar{0}_c\rangle+c_2|\bar{1}_c\rangle$, cf. Eq.~(\ref{eq-psi-c}) in the main text, the code state collapses into a state with a certain parity after each measurement, with probability $P_{m}(t)=\text{Tr}[\hat{\Pi}_{m}\hat{\rho}(t)\hat{\Pi}_{m}^{\dag}]$. 
Regardless of whether a single-photon loss error occurs or not, we simply update our knowledge of the state and continue the dynamical process.
We divide the unit interval $[0, 1]$ into two sections with lengths $P_{0}$ and $P_{1}$, respectively. The target state is updated by
generating a random number $\varepsilon_{r}\in[0, 1]$ and determining its position within the unit interval. Specifically, we perform the projection 
$\hat{\rho}(t)\rightarrow{\hat{\Pi}_{m}\hat{\rho}(t)\hat{\Pi}_{m}}/{P_{m}(t)}$,
cf. Eq.~(\ref{project1}) in the main text,
if the random number falls in the $m$-th section. Correspondingly, we calculate the fidelity of the projected code state by
$F_c(t)\rightarrow\text{Tr}\Big[\sqrt{\hat{\rho}_{T}^{1/2}\hat{\rho}(t)\hat{\rho}_{T}^{1/2}}\Big]$,
cf. Eq.~(\ref{upF}) in the main text,
where $\hat{\rho}_{T}=|\psi_T\rangle\langle\psi_T|$ is the target state updated according to,
\bea\label{}
|\psi_T\rangle&=&c_1|\bar{0}_c\rangle+c_2|\bar{1}_c\rangle\rightarrow c_1|\bar{1}_e\rangle+c_2|\bar{0}_e\rangle\rightarrow c_1|\bar{1}_c\rangle+c_2|\bar{0}_c\rangle\rightarrow c_1|\bar{0}_e\rangle+c_2|\bar{1}_e\rangle\rightarrow c_1|\bar{0}_c\rangle+c_2|\bar{1}_c\rangle,
\eea
cf. Eq.~(\ref{eq-update}) in the main text. It is worthy to point out that the above state updating procedure, after each single-photon loss, is $\textit{not}$ equal to applying $\hat{\rho}_T\rightarrow\hat{a}\hat{\rho}_T\hat{a}^\dagger/\mathrm{Tr}(\hat{a}\hat{\rho}_T\hat{a}^\dagger)$. This is because the normalization factors of the single-photon corrupted cat code states ($\hat{a}|\bar{0}_c\rangle,\ \hat{a}|\bar{1}_c\rangle$) and error states ($\hat{a}|\bar{0}_e\rangle,\ \hat{a}|\bar{1}_e\rangle$) are generally not the same. Such a discrepancy in normalization is also one of the errors resulting in the slowly increasing infidelity after each measurement, cf. Fig.~\figpanel{Fig-AutoQEC}{a} in the main text.

\twocolumngrid


%

\end{document}